\documentclass[prx,twocolumn,amsmath,amssymb,floatfix,footinbib,bibnotes,longbibliography]{revtex4-1}

\pdfoutput=1

\usepackage[colorlinks=true,citecolor=blue,linkcolor=magenta]{hyperref}
\usepackage{amsmath}
\usepackage{graphicx}
\usepackage{dsfont}
\usepackage{changepage}
\usepackage{fancyhdr}
\usepackage{amsthm, amssymb}
\usepackage{float}
\usepackage{array}
\usepackage[usenames,dvipsnames]{color}
\usepackage{slashed}

\newcommand{\be}{\begin{align}}
\newcommand{\ee}{\end{align}}
\def \tr{{\mbox{tr~}}}

\def \be{\begin{equation}}
\def \ee{\end{equation}}
\def \ba{\begin{array}}
\def \ea{\end{array}}
\def \bea{\begin{eqnarray}}
\def \eea{\end{eqnarray}}
\def \nn{\nonumber}

\def \bu{{\bf u}}

\def \e{{\epsilon}}
\def \ve{{\varepsilon}}
\def \l{{\lambda}}
\def \L{{\Lambda}}
\def \a{{\alpha}}

\def \b{{\beta}}
\def \g{{\gamma}}

\def \d{{\delta}}
\def \w{{\omega}}
\def \s{{\sigma}}

\def \k{{\kappa}}

\def \e{{\epsilon}}

\def \G{{\Gamma}}

\def \ba{\begin{align*}}
\def \ea{\end{align*}}

\newcounter{indice}

\def \mrm{\mathrm}

\def \bs{\boldsymbol}
\def \mc{\mathcal}

\begin{document}

\title{  Superconductivity near a Ferroelectric Quantum Critical Point in Ultralow-Density Dirac Materials   }

\author{ Vladyslav Kozii$^1$, Zhen Bi$^1$, and Jonathan Ruhman$^{2}$}
\email{All authors contributed equally.}
\affiliation{ {
1. Department of Physics, Massachusetts Institute of Technology, Cambridge, MA 02139, USA  \\
2. Department of Physics, Bar Ilan University, Ramat Gan 5290002, Israel  }}
\begin{abstract}
{ The experimental observation of superconductivity in doped semimetals and semiconductors, where the Fermi energy is comparable to or smaller than the characteristic phonon frequencies, is not captured by the conventional theory.
In this paper, we propose a mechanism for superconductivity in ultralow-density  three-dimensional Dirac materials based on the proximity to a ferroelectric quantum critical point. We derive a low-energy theory that takes into account both the strong Coulomb interaction and the direct coupling between the electrons and the soft phonon modes. We show that the Coulomb repulsion is strongly screened by the lattice polarization near the critical point even in the case of vanishing carrier density.
Using a renormalization group analysis, we demonstrate that the effective electron-electron interaction is dominantly mediated by the transverse phonon mode. We find that the system generically flows towards strong electron-phonon coupling.
Hence, we propose a new mechanism to simultaneously produce an attractive interaction and suppress strong Coulomb repulsion, which does not require retardation. For comparison, we perform same analysis for covalent crystals, where lattice polarization is negligible. We obtain qualitatively similar results, though the screening of the Coulomb repulsion is much weaker. We then apply our results to study superconductivity in the low-density limit. We find strong enhancement of the transition temperature upon approaching the quantum critical point. Finally, we also discuss scenarios to realize a topological $p$-wave superconducting state in covalent crystals close to the critical point.
 }
\end{abstract}
\maketitle

\tableofcontents

\section{Introduction}

A key step in the formation of a superconductor is the pairing between electrons. In spite of the strong Coulomb repulsion in free space, at low energy electrons experience an effective attraction in the presence of a lattice.
Thus, superconductivity essentially relies on a mechanism that simultaneously reduces the Coulomb repulsion and generates a strong attractive interaction.

In  simple (elemental) metals, such an attraction originates from the interchange of longitudinal phonons, which couple to the electronic density.
To allow for this attraction to overcome the Coulomb repulsion, however, it is essential that the crystal vibrations are much slower than electronic motion. In terms of energy scales, this requirement implies that the Fermi energy is much larger than the Debye frequency. In the intermediate frequency regime, between these two scales, the Coulomb repulsion is logarithmically suppressed, while the phonon interaction is unaffected \cite{Tolmachev1958,Bogoliubov1958,Morel1962}. As a result, the net interaction between electrons may become attractive below the Debye energy.

From this perspective, systems of low carrier concentration, such as doped semimetals and semiconductors, are not expected to exhibit superconductivity. First, {\it they have a low Fermi energy, which is comparable to, or even smaller than, the typical Debye frequency}, and, thus, does not allow for the dynamical screening of the repulsion. Moreover, the superconducting transition temperature is exponentially sensitive to the {\it density of states, which is typically two orders of magnitude smaller in doped semimetals and semiconductors compared to standard metals}.  Thus, naively, attainable transition temperatures require an unphysically large interaction strength.

Surprisingly, however, superconductivity in doped semimetals and semiconductors is ubiquitous. It was first discovered in SrTiO$_3$~\cite{Schooly1964} and later in many other materials~\cite{Bustarret2008}. To the best of our knowledge~\footnote{Superconductivity has been measured in Zr-doped SrTiO$_3$ where the density was argued to be even lower~\cite{eagles2016comment}}, the lowest-density superconductors discovered to date are Tl-doped PbTe \cite{Matsushita2006}, Sr-doped Bi$_2$Se$_3$ \cite{Liu2015}, YPtBi~\cite{Butch2011}, SrTiO$_{3-x}$~\cite{Lin2014,bretz2019superconductivity}, and elemental bismuth~\cite{Prakash2017}.
It is noteworthy that, except for SrTiO$_3$, all of these materials are either narrow-band topological insulators or topological semimetals. The common feature they share is a near crossing of their  conduction and valence bands, for example, such as in Dirac material.  These experiments, thus, impose two theoretical challenges: (1) How is the Coulomb repulsion screened when the dynamical timescale of the pairing interaction is comparable to, or even smaller than, the electronic timescale (the so-called antiadiabatic limit)? (2) What is the source of attraction, which is strong enough to deal with such a small density of states?

A variety of theoretical frameworks have been proposed to discuss superconductivity in the limit of low density, including  polar phonons~\cite{Gurevich1962,Takada1980,Savary2017,rowley2018superconductivity,gastiasoro2019phonon}, plasmons~\cite{Takada1978,Takada1980,Ruhman2016,Ruhman2017b}, multi-band effects~\cite{Koonce1967,Binnig1980}, soft optical phonons~\cite{Appel1969}, the charge Kondo effect~\cite{Matsushita2005}, and instantaneous attraction~\cite{Eagels1969,Gorkov2017}. It is particularly important to single out the seminal contribution of the authors of Ref. \cite{Gurevich1962}, who pointed out an essential ingredient in any theory of low-density superconductivity: a long-ranged attractive interaction.  When the range of the attractive interaction is comparable to the distance between conduction electrons, it naturally competes with the small density of states and thus paves the way to solve the second theoretical challenge we specified above.
This effect is similar to the phenomenon of Wigner crystallization, where the long-ranged Coulomb  interaction dominates the kinetic energy in the dilute limit rather than at high density. Thus, the constraint of long-ranged attraction narrows down the range of viable pairing mechanisms in the extreme low-density limit. Such interaction may result from a dynamically screened Coulomb repulsion~\cite{Gurevich1962,Takada1978}, fluctuations of an order parameter close to a quantum critical point~\cite{Chubukov1,Lederer2015,Metlitski,Chubukov2},  spin fluctuations~\cite{Scalapino1999,Hirschfeld_2011}, or Goldstone mode fluctuations in certain types of spontaneously broken continuous symmetries~\cite{Watanabe16314}.

\begin{figure}
 \begin{center}
    \includegraphics[width=1\linewidth]{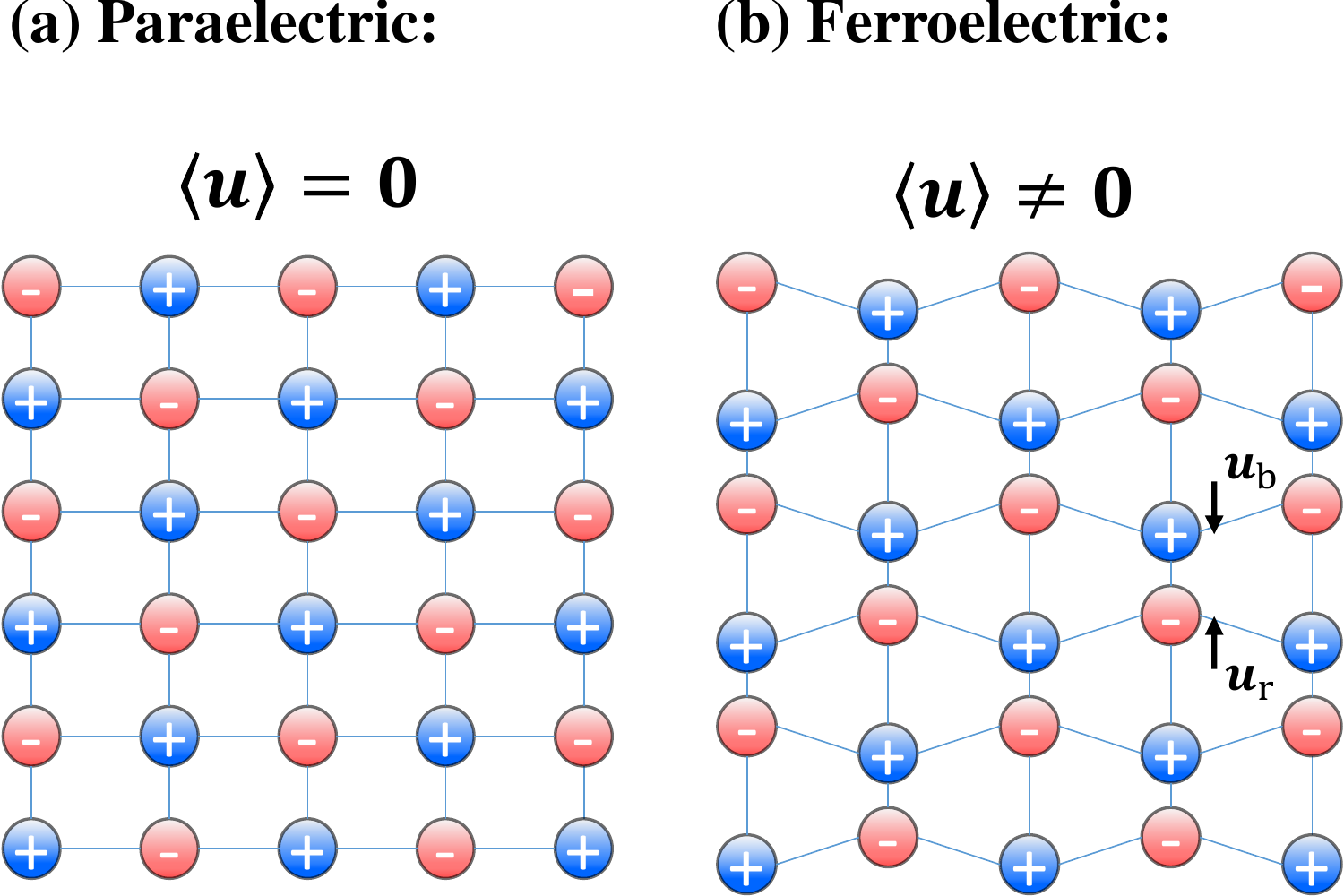}
 \end{center}
\caption{(a) A diatomic cubic ionic crystal in the paraelectric phase. The two ions are represented by blue (+) and red (-) circles (e.g., blue for Pb and red for Te). Unless these two atoms are identical, they will have an average charge imbalance.
(b) The ferroelectric phase (inversion breaking) is characterized by a uniform optical phonon displacement vector $\bs u = \bs u_b - \bs u_r$. Because of the charge imbalance between the ions, this phase is also characterized by a finite dipolar polarization density $\bs P = Q \bs u$, where $Q$ is the charge.    }
 \label{fig:dist}
\end{figure}

It is interesting to notice that SrTiO$_3$, PbTe, and SnTe naturally reside close to a paraelectric (PE)-ferroelectric (FE) phase transition~\cite{Rowley2014,Jantsch2001}, which can be tuned in various manners.
Indeed, it has been proposed theoretically that the superconducting state in low density SrTiO$_{3-x}$ results from a pairing interaction mediated by FE fluctuations near the quantum critical point (QCP)~\cite{Edge2015}.
Following this proposal, recent experiments reported enhancement of $T_c$ in the vicinity of the QCP~\cite{Stucky2016,Rischau2017,rowley2018superconductivity,tomioka2019enhanced}.
This proposal has also led to a number of theoretical studies discussing different aspects of the problem~\cite{Wolfle2018,Kedem,Kanasugi2018,Arce-Gamboa2018}.
However, an important question that remains open is the microscopic origin of a strong electronic coupling to the soft FE fluctuations, which is na\"{i}vely expected to be weak~\cite{evrard1972polarons,AMbook,mahan2013many} (we also refer to a recent comment on the subject~\cite{ruhman-comment} and the recent preprint~\cite{van2019possible}).

To explain the origin of the common belief concerning weak coupling between electrons and soft FE modes in low-density ionic crystals, we briefly review some basic facts regarding the FE QCP. The PE-FE transition is essentially a structural transition where the order parameter is a vector (i.e., a lattice distortion), which spontaneously breaks inversion and rotation symmetries in the ordered state. As an example, we consider the diatomic ionic crystal in Fig.~\ref{fig:dist}. In the FE phase, the two ions in the unit cell are distorted from their cubic Bravais lattice positions. Because of a charge imbalance, the ions induce a uniform electric polarization density. Thus, the transition is described by a soft optical phonon mode associated with the relative displacement of the two charged ions. This phonon mode has three polarizations: one longitudinal optical (LO) and two transverse optical (TO). The long-range dipolar interactions between lattice distortions, however, prevent the LO mode from softening near the transition~\cite{khmelnit1971low,strukov2012ferroelectric,kvyatkovskii2001quantum,roussev2001quantum,Rowley2014}. Consequently, the soft bosonic modes associated with the FE QCP are purely transverse.
On the other hand, the simplest coupling between conduction electrons' density and lattice is Fr\"{o}hlich coupling, which only involves longitudinal modes~\cite{frohlich1950xx,evrard1972polarons,mahan2013many}. As a result, the interaction between electrons and soft FE modes is typically considered as weak.

Despite the above discussion, the direct coupling of gapless electrons to soft transverse phonon (ferroelectric) modes is possible in  multiorbital systems~\cite{MAHAN1965751,SHAPIRO1988239}.  For time-reversal symmetric systems with a single Fermi surface multi-orbital effects can only manifest themselves due to the presence of spin-orbit-coupling ~\cite{Fu2015,Kozii2015}. In this paper, we use this idea to explicitly derive a complete low-energy theory capturing all gapless degrees of freedom at the FE QCP in three-dimensional Dirac materials, a manifestly multi-orbital system with strong spin-orbit coupling, and show how transverse modes couple to gapless electrons in the long-wavelength limit. We use the renormalization group (RG) approach to study the tendency of this theory towards strong coupling close to the critical point.  We find that the proximity to the FE QCP in the low-density limit leads to strong screening of the Coulomb repulsion between electrons by the crystal. Concomitantly, the interband transitions across the Dirac point enhance the effective coupling of electrons to the soft TO phonon modes.

%
%
Thus, the attraction mediated by TO modes generically overcomes the Coulomb repulsion close to a FE QCP. Interestingly, this result is valid even at vanishing electronic density, when the effects of Coulomb repulsion are expected to be strong. Therefore, our theory is distinct from the standard Tolmachev-Anderson-Morel mechanism~\cite{Tolmachev1958,Bogoliubov1958,Morel1962}, since it does not require  the phonon frequency to be smaller than the Fermi energy in order to produce net attraction at zero frequency. Finally, we analyze the possible superconducting instabilities from the interaction mediated by the critical phonon mode. We find strong  enhancement of the transition temperature $T_c$ due to the enhancement of the electron-phonon coupling close to the critical point.

For completeness and comparison, we perform similar analysis for covalent  (non-ionic) crystals, where both LO and TO modes are soft at the critical point. In this case, negligible lattice polarization does not allow for a spontaneous dipolar moment of the crystal, so a ``ferroelectric'' phase is simply characterized by broken inversion symmetry. Because of the lack of lattice polarization, the screening of the Coulomb repulsion is only due to interband transitions; consequently, it is much weaker  than in ionic crystals. Namely, it is logarithmic, similar to the case without soft phonons~\cite{Hosur,isobe2012}.  We find that the dimensionless coupling constant associated with the coupling to the longitudinal mode also flows logarithmically to zero, while the coupling to transverse modes remains relevant, as before.
Thus, this result is an example where attraction may overcome repulsion without any requirement on the Fermi energy, even without screening from the crystal. We point out that this result is potentially relevant to other nonpolar critical modes that couple to Dirac points.
We also find that the interplay between phonon-mediated attraction and the Coulomb repulsion in these covalent crystals opens a possibility of topological $p$-wave superconductivity in a certain range of parameters.

Beside the fundamental theoretical importance, our study of superconductivity from ferroelectric quantum critical fluctuations is also motivated by a realistic system:  the ionic alloy Pb$_{1-z}$Sn$_z$Te, which undergoes a FE phase transition at $z=z_{FE}\approx 0.25$~\cite{Jantsch2001}. When $z$ is further increased above $z_T \approx 0.41$, the alloy undergoes a second, topological phase transition, between a trivial insulator and a topological crystalline insulator~\cite{hsieh2012topological}. The topological transition entails gapless Weyl points close to the $L$-points of the Brillouin zone~\cite{Liang2017}. When doped with Tl or In atoms, this alloy becomes superconducting, and the transition temperature exhibits a peak at some intermediate value of $z$~\cite{Parfen2001}. While doped  Pb$_{1-z}$Sn$_z$Te seems a promising candidate for our theory, a word of caution is needed, since certain features require further understanding. In pure PbTe, for example, superconductivity appears only when doped with Tl. Additionally, it has been found that the superconducting state emerges only above a critical density, where additional electron pockets become populated~\cite{Giraldo-Gallo2018}.

The remainder of the paper is organized as follows.  We first summarize our main results while providing intuitive pictures in Sec.~\ref{Sec:summary}. In Sec.~\ref{sec:model},  we present a complete low-energy theory for Dirac materials near a FE QCP. In Sec.~\ref{sec:RG}, we use the RG approach to demonstrate that the Coulomb repulsion is strongly screened by lattice polarization, while the coupling between electrons and soft transverse phonons is significantly enhanced. We find qualitatively similar results for covalent crystals, though the screening of Coulomb repulsion in this case is much weaker because of negligible lattice polarization. Finally, in Sec.~\ref{sec:SC}, we analyze the possible superconducting instabilities from the interaction mediated by the critical phonon (ferroelectric) mode. We find strong  enhancement of the transition temperature $T_c$ due to the enhancement of the electron-phonon coupling close to the critical point. Additionally, we discuss scenarios for $p$-wave superconductivity  originating from the interplay between phonon-mediated attraction and Coulomb repulsion in covalent crystals. We finish with a short summary and discussion in Sec.~\ref{Sec:conclusions}.

\section{Summary of main results}\label{Sec:summary}

Before moving on to the main part of the paper, where we carefully analyze the FE QCP in a Dirac system using the RG technique, we first present our main results at a non-technical level. We aim to qualitatively explain how our findings deal with the long standing challenge of understanding superconductivity in low-density systems.

As explained in the introduction, the observation of superconductivity in extremely dilute metals poses two questions which are the main focus of this paper:
\begin{itemize}
    \item[(1)]{How is the Coulomb repulsion screened in a system with relatively small Fermi energy (the so-called antiadiabatic limit)?}
    \item[(2)]{What is the source for strong attraction that can overcome the small density of states in these systems?}
\end{itemize}

\begin{figure}[t!]
 \begin{center}
    \includegraphics[width=1\linewidth]{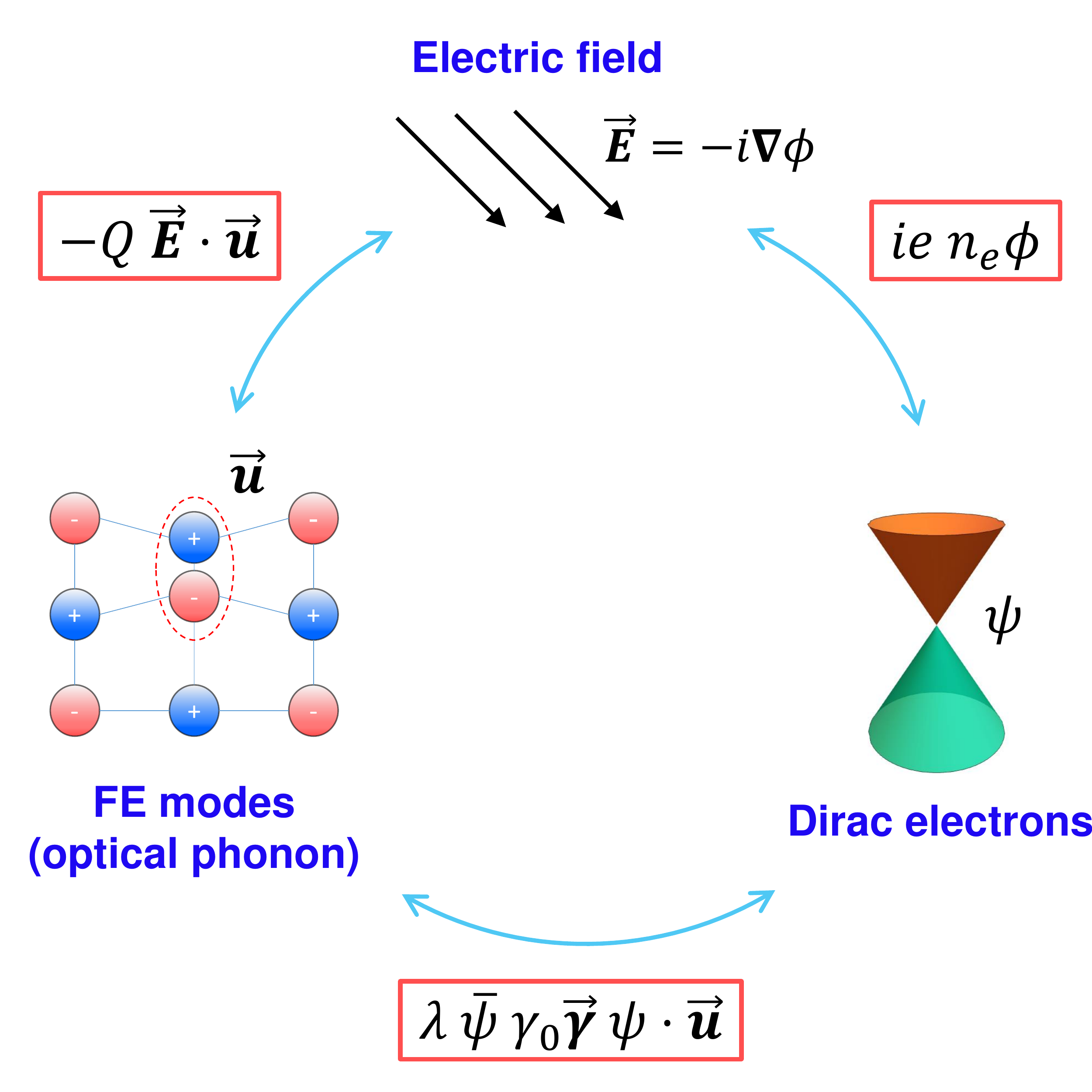}
 \end{center}
\caption{The three fields participating in the low-energy effective field theory Eq.~(\ref{S_fixed}) and the couplings between them. The fields are the ferroelectric fluctuations (optical phonon distortions), Dirac electrons and {\it static} electric field. The coupling of ferroelectric mode to the electric field leads to the gapping out the longitudinal optical mode (LO-TO splitting) and to the well-known Fr\"ohlich coupling, see Eq.~\eqref{Eq:Frohlich}. The coupling of the electrons to the electric field generates the Coulomb repulsion. An important new element, which is one of the key points of the paper, is the direct coupling $\lambda$ between the optical phonon distortions and the Dirac electrons, see Eq.~\eqref{Spsi-u}.  }
 \label{fig:couplings}
\end{figure}

To answer these questions, we focus on a concrete model consisting of three essential ingredients, which are all mutually coupled (see Fig.~\ref{fig:couplings}). The ingredients are: Fluctuations of the FE order parameter close to a FE QCP (optical phonon distortions), gapless Dirac fermions,
and the static electric field~\footnote{Note that here we neglect the fast dynamics of the electromagnetic gauge field and, as a consequence, the electric field is purely longitudinal  (i.e., it derives from a potential).}.

The key new element that plays a crucial role in our study is the direct coupling $\lambda$ between electrons and optical phonons  (especially the TO modes). The coupling is allowed in materials with strong spin-orbit coupling such as Dirac semimetals. The intuitive picture behind this coupling is the following: When the unit cell distorts due to a FE optical phonon, inversion symmetry is locally broken. In the presence of spin-orbit coupling, this process induces a Rashba effect modifying the electronic dispersion, thus coupling electrons to phonons.

\begin{figure}[t!]
 \begin{center}
    \includegraphics[width=1\linewidth]{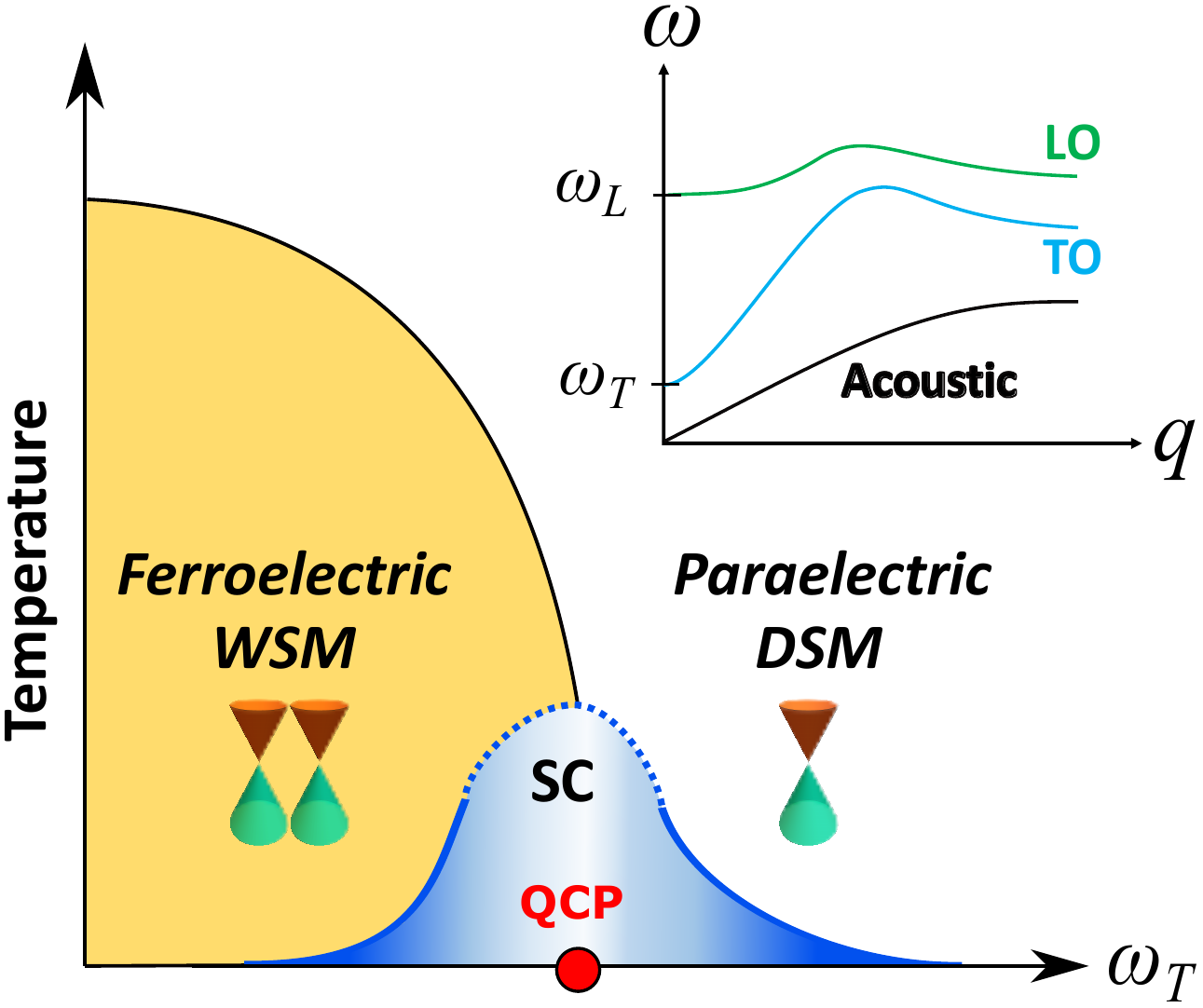}
 \end{center}
\caption{ Schematic phase diagram of the 3D Dirac semimetal (DSM) near a ferroelectric (FE) quantum critical point (QCP). The FE QCP separates the inversion-symmetric DSM and inversion-broken Weyl semimetal (WSM) phases. Upon approaching the phase transition (controlled by $\omega_T$), there is a region where we predict superconductivity at ultralow density (blue dome). Due to the long-ranged dipolar forces between longitudinal optical (LO) modes, close to the FE QCP, only the transverse optical (TO) modes soften, as shown  in the inset (the LO-TO splitting). We consider the  direct coupling of these TO modes to the Dirac fermions through spin-orbital effects. RG analysis shows that this coupling is a relevant perturbation that renders the Dirac point unstable at low energy. We find that this type of QCP combines two  ingredients which are essential for superconductivity at low density: (1) Strong attraction that compensates small density of states and (2) screening of the Coulomb repulsion between electrons by the LO modes. The dashed line denotes the region of strong coupling in the immediate vicinity of the critical point where our weak-coupling analysis breaks down. }
 \label{fig:schematic_phase_diagarm}
\end{figure}

Interestingly, the first question raised above, screening of the Coulomb repulsion in the antiadiabatic limit, is naturally dealt with in this model. In the case of ionic crystals,  the polar phonons provide the screening. The important prerequisite, however, is the proximity of the system to a FE QCP. Indeed, according to the  Lyddane-Sachs-Teller relation~\cite{lyddane1941polar}, the low-energy dielectric constant behaves as $\ve_0 \sim \w_L^2 / \w_T^2$~\cite{lyddane1941polar}. At the FE QCP, the TO phonon frequency vanishes, while the LO phonon remains gapped due to its polar nature (see the inset in Fig.~\ref{fig:schematic_phase_diagarm} for a schematic dispersion). Consequently, $\omega_T\to 0$, while $\omega_L$ remains finite near the FE QCP, so $\ve_0$ diverges, leading to the suppression of the long-ranged Coulomb interaction. Additioinally, we find that even in the absence of crystal screening (like in covalent crystals that we discuss below), the interband transitions across the Dirac dispersion also effectively screen the Coulomb repulsion in the antiadiabatic limit.

To deal with the second question about the source of the sufficiently strong attraction between electrons, we recall the seminal result by Gurevich et al.~\cite{Gurevich1962}, who pointed out that the low-density SC necessarily requires long-ranged interaction. In our theory, there are two key elements that provide such interaction. First, near the FE QCP, we have soft (nearly gapless) TO phonon mode. This fact itself, however, does not guarantee the long-ranged interaction; otherwise, the acoustic phonons would be sufficient for this purpose. The crucial element of our theory that allows for such long-ranged interaction is exactly the direct coupling between electrons and optical phonons $\lambda$ that we discussed above. Indeed, this coupling does not vanish at small momentum transfer and is generically present in 3D Dirac materials. Then, it is straightforward to show that the effective attraction between electrons mediated by TO phonons is sufficiently long-ranged, which compensates the low density of states. Putting all the described key points together, we conclude that the combination of a finite electron-phonon coupling and the proximity to the FE QCP is sufficient for superconductivity in the low-density 3D Dirac materials. This result is schematically summarized in Fig.~\ref{fig:schematic_phase_diagarm}.

To make the  above arguments more quantitative and rigorous, we study our model using the renormalization group technique. In particular, we find that, in the weak-coupling regime, the dimensionless coupling constant between electrons and TO phonons, $\b$, is marginally relevant [see Eqs.~(\ref{RG_eqs_polar}) and (\ref{Eq:beta*})] and, hence, gets enhanced at small energies. The main origin of this enhancement is the reduction of the TO phonon velocity due to a cloud of virtual particle-hole excitations that is generated as it propagates through the crystal.
The velocity reduction increases the TO phonon density of states at low energies, which enhances the resulting phonon-mediated interaction. The particle-hole excitations, which are responsible for the velocity reduction, are of the interband type (i.e., across the Dirac node). Therefore, the nearly touching between the conduction and valence bands, characterizing the Dirac dispersion, is important for the enhancement of $\b$. This enhancement, in turn, leads to a sufficient increase in the superconducting transition temperature $T_c$, see Eqs.~(\ref{Eq:Tc0})-(\ref{Eq:Tcenhanced}).

As a final step, we calculate the superconducting instabilities with the renormalized parameters. We work within the weak-coupling approximation, and assume small but finite Fermi surface with the radius $k_F$. We find that the strongest superconducting instability is in the $s$-wave channel, but the vector-type $p$-wave also exhibits weaker instability towards the Cooper pairing. For these two channels, we compute the corresponding transition temperatures, see Eqs.~\eqref{Eq:Tc0}-\eqref{Eq:Tcp}. As is usual for weak-coupling approximation, $T_c$ is exponentially sensitive to the renormalized dimensionless coupling $\b^*$ and the square of the ratio $k_F/\L_r$, where $\L_r$ can be viewed as the inverse correlation length of the FE order parameter or, equivalently, the range of the TO phonon-mediated attraction between electrons. Since we found $\beta$ to be a relevant operator, both $\b^*$ and $k_F/\L_r$ can be made of the order of 1 near the FE QCP, thus leading to a significantly large transition temperature despite low density of states. Our rough estimates show that one can reach $T_c$ of the order of several Kelvin sufficiently close to the critical point even at densities $n_0\sim 10^{17}$cm$^{-3}$.

The above discussion is relevant for ionic crystals. For completeness, we also consider the case of covalent crystals, where the atoms in Fig.~\ref{fig:dist} would be neutral, and thus, the optical phonons would be decoupled from the electric field. In this case, the LO mode is also gapless at the transition point, and, consequently, it no longer screens the Coulomb interaction.  Our RG analysis shows that the same interband processes that enhance the coupling $\b$ suppress the fine-structure constant $\a$ and the coupling to the LO phonon $\tilde \b$, see Eqs.~(\ref{RG_eqs_non_polar2})-(\ref{Eq:betatilde}). The former is simply the interband screening of the Coulomb interaction~\cite{Hosur}. Consequently, the end result is qualitatively the same: At the critical point, the TO phonon mediated interaction is the dominant one, and there is still a mechanism to screen the Coulomb repulsion, which does not rely on retardation (in this case, it is purely electronic). The crucial difference, however, is that the screening of the Coulomb interaction is only logarithmic and, consequently,  weaker than in the case of ionic crystals.  The fact that crystal screening is not necessary to suppress  the Coulomb repulsion at scales higher than the Fermi level also implies that our analysis for the FE QCP can be extended to other types of critical points in Dirac semimetals. The fact that almost all known low-density superconductors carry semimetallic features in their band structure raises the question of whether such an antiadiabatic screening of the Coulomb repulsion plays an important role in the emergence of superconductivity in these systems.

\section{Model}
\label{sec:model}
We now turn to the main part of the paper.
{We first consider the low-energy effective field theory of a Dirac semimetal near the ferroelectric transition. The Euclidean (imaginary time) action is given by the sum
\be \mc S = \mc S_{\psi}+\mc S_{u}+\mc S_{\phi}+\mc S_{\psi u}+\mc S_{\psi \phi}+\mc S_{u \phi},\label{S}\ee
where the first three terms describe the dynamics of the fermions $\psi$, the optical phonon field $u$, and the Coulomb field $\phi$, respectively, while the latter three describe their interactions. The Coulomb field $\phi$ should be considered as a Hubbard-Stratonovich decomposition of the instantaneous Coulomb interaction. Now we specify these terms in detail.}

\subsection{Quadratic terms}
\noindent
{\it The electron term --} The electron quadratic term (motivated by the model of the PbTe crystal~\cite{hsieh2012topological}) reads
\be
\mc S_{\psi} = \sum_{n=1}^N\int d^4 x \,\bar\psi_n \left(\gamma_0 \partial_0+v_F \gamma_j \partial_j  +m -  \gamma_0 \ve_F\right) \psi_n, \label{Spsi}
\ee
where $\psi_n$ is a four-component Dirac spinor, $n = 1,\ldots,N$ denotes different fermionic flavors (number of Dirac nodes), summation over $j=x,\,y,\,z$ is implied ($\partial_0$ stands for derivative in imaginary time), and $\bar \psi_n \equiv \psi_n^\dag \g^0$. Parameters $v_F,$ $m$, and $\ve_F$ stand for electron velocity, Dirac mass, and Fermi energy, respectively. We use Hermitian gamma matrices $\{\g_{0}, \g_{1},\g_{2},\g_{3} \} = \{  \s^1\otimes \s^0,\s^2\otimes \s^1,\s^2\otimes \s^2,\s^2\otimes \s^3 \}$ and $\g_5 = \g_0\g_1 \g_2\g_3$, where $\sigma_i$ are usual Pauli matrices. Notice that here we have assumed an isotropic dispersion by taking the same velocity~$v_F$ in all directions. The anisotropic case does not modify the main qualitative results of this paper, and therefore we comment on it in Appendix~\ref{App:Anisotropy}. For generality, we have assumed a nonzero mass term $m$ and a finite Fermi energy~$\ve_F$. However, we will neglect them in our RG analysis, assuming that they are much smaller than other relevant energy scales.

We note two important discrete symmetries of Eq.~(\ref{Spsi}): inversion symmetry $\mc P$ and time-reversal symmetry $\mc T$. The action of these symmetries in terms of Dirac matrices is given by $\mc P  = \g_0$ and $\mc T = \g_1\g_3 K$, respectively, where $K$ is complex conjugation.

\noindent
{\it The phonon term --} Next, we consider the dynamics of the phonon modes, which become soft at the FE phase transition. To have an intuitive picture in mind, we consider the scenario in which the FE order is dominantly generated by a lattice distortion. For simplicity, we consider a cubic ionic crystal with two atoms in the unit cell (the rocksalt structure of the IV-VI semiconductors, see Fig.~\ref{fig:dist}).
We label the two sublattices by $b$ and $r$ corresponding to the ``blue" and ``red" ions, respectively, which have equal in magnitude and {\it opposite} sign charges.
Each sublattice has a corresponding phonon displacement field $\bs u_r$ and $\bs u_b$.
As usual, there are two modes: a gapless acoustic mode given by the sum $\bs u_{ac} = (\bs u_r + \bs u_b)/2$ and a gapped optical branch given by the difference $\bs u = \bs u_r - \bs u_b$.
Near the FE transition, the acoustic mode is irrelevant, while the optical branch becomes nearly gapless and is described by the effective action
\begin{multline}
\mc S_u =\int d^4 x {1 \over 2}\;u_j[\left(-\partial_0^2  + {  \w_T^2 }\right)\d_{jl}\label{Su}\\
-  c_T^2\left(\nabla^2\d_{jl}-\partial_j\partial_l\right)-c_L^2\partial_j\partial_l ]u_l + V (u_j u_j)^2.
\end{multline}
Here $c_L$ and $c_T$ are the longitudinal and transverse phonon velocities, respectively, $\w_T$ is the phonon mass, which is the tuning parameter to the transition, and
$V$ is the lowest-order symmetry allowed anharmonic correction to the phonon energy (where we have neglected additional anisotropic terms allowed by the cubic symmetry~\cite{khmelnit1971low,strukov2012ferroelectric,kvyatkovskii2001quantum,roussev2001quantum}). Again, summation over $j,\,l = x,\,y,\,z$ is implied.

\noindent
{\it The Coulomb term --}
The third quadratic term describes the Coulomb potential:
\be
\mc S_\phi ={\ve_{\infty} \over 8\pi } \int d^4 x \, \left(\nabla\phi \right)^2,  \label{Sphi}
\ee
where $\ve_{\infty}$ is the bare dielectric constant,  which accounts for the contribution of core electrons. This contribution is due to the transitions between the high-energy atomic configurations, and does not include the contributions from the lattice dynamics or electronic interband transitions close to the Dirac point.

\subsection{Coupling terms}
We now consider the couplings between the fields introduced in Eqs.~\eqref{Spsi}-\eqref{Sphi}.

\noindent
{\it Electron-Coulomb coupling --}  We start with the coupling between the Dirac electrons and the Coulomb potential
\be
\mc S_{\psi \phi} = i\,e\, \int d^4 x \,\rho_e \,\phi, \label{Spsi-phi}
\ee
where $\rho_e = \sum_n \bar \psi_n \g_0 \psi_n$ is the electronic density.

\noindent
{\it Phonon-Coulomb coupling --}
The coupling of the ferroelectric phonon modes to the Coulomb potential follows from Eq.~\eqref{Spsi-phi} by noting that the deviations of the ``red" and ``blue" ionic density from  the average equilibrium value $\rho_0$ in the long-wavelength limit are given by $ \rho_r = \rho_0(1-\nabla \bs u_r)$ and $ \rho_b = \rho_0(1-\nabla \bs u_b)$, respectively. Given that the ionic charges are of equal magnitude and opposite signs, the coupling of the lattice to the Coulomb field $\phi$ is given by
\be
\mc S_{u\phi} = i\,Q \,\int d^4 x\left( \rho_r -  \rho_b \right)\phi  = i\,Q\rho_0 \, \int d^4 x\, \nabla \bs u\,\phi, \label{Su-phi}
\ee
where $Q$ is the ionic charge on ``blue" sites (charge on ``red" sites equals $-Q$). For the purpose of further analysis, it is convenient to absorb factor $\rho_0$ by redefining $Q \rho_0 \to Q.$

Notice that the form of the coupling \eqref{Su-phi} implies that only the {\it longitudinal} phonon mode couples to the Coulomb field.
We also point out that after integrating Eq.~\eqref{Su-phi} by parts one gets a dot product between the polarization density $\bs P = Q \bs u$ and the electric field $\bs E =- i \nabla \phi $. Therefore, this equation can also be viewed as the action of a dipole moment density in an electric field. Finally, in the case of a nonpolar covalent crystal (e.g., elemental bismuth), all atoms in the unit cell are neutral, leading to a vanishing coupling $Q = 0$.

\noindent
{\it Electron-Phonon coupling --}
We now consider the coupling between the Dirac electrons and the phonon modes.
We write down this coupling from general symmetry arguments. The phonon mode $\bs u$ is a time-reversal invariant vector. Inspecting all possible {\it local} Dirac bilinears specified in Table~\ref{tab:gamma}, we find that the only Dirac bilinear that forms a time-reversal symmetric vector and, thus, is allowed to couple to the phonon displacement field is $\bar \psi \g_0 \g_j \psi$. Therefore, the corresponding coupling is given by
\be
\mc S_{\psi u} = \l \sum_{n=1}^N\int d^4 x \, \bar \psi_n\, \g_0 \g_j \,\psi_n \,u_j. \label{Spsi-u}
\ee
The microscopic origin of this coupling is a combination of the interorbital hybridization induced by the lattice distortion $\bs u$ and spin-orbit coupling. This effect is similar to the Rashba effect, which arises when inversion is broken in a system with spin-orbit coupling. The analogy is made by noting that the optical phonon distortion is essentially an inversion breaking field. It is also worth noting that the form of the coupling Eq.~\eqref{Spsi-u} can be derived from the action of inversion breaking on a Dirac node~\cite{bzduvsek2015weyl}.  To estimate the magnitude of the coupling accurately, however, an {\it ab-initio} calculation is required.   Finally, we also note that  in Eq.~\eqref{Spsi} we assumed that the Dirac cones occur at the inversion symmetric points in the Brillouin zone. In the case they do not, the coupling can also include inter-flavor scattering.

We emphasize that, for simplicity, we consider a rotationally symmetric model in the main text. We discuss the possible effects of the cubic anisotropy in Appendix~\ref{App:Anisotropy}.

\begin {table*}
\caption { Parity and time-reversal symmetry of the 16 ($k$-independent) Dirac bilinears. } \label{tab:gamma}
\begin{center}
    \begin{tabular}{| c | c | c | c | c |}
    \hline
     Bilinear & $ 1,\g_0 $    & $ \{i\g_1,i\g_2,i\g_3\} ,i\g_5$ & $\{\g_0 \g_1,\g_0 \g_2,\g_0 \g_3\}, \g_0 \g_5 $ &$\{ i\g_0\g_1 \g_5,i\g_0\g_2 \g_5,i\g_0\g_3 \g_5,i\g_0\g_1 \g_2,i\g_0\g_1 \g_3,i\g_0\g_2 \g_3\} $ \\ \hline
    $\mc P = \g_0$ &                         $+$  & $-$ &$ - $ &$+$  \\ \hline
$\mc T = \g_0\g_2\g_5 K$ &$+ $ &$-$ &$+$ &$ -$ \\ \hline
    \end{tabular}
    \end{center}
    \end {table*}

\section{Renormalization group analysis near the critical point}
\label{sec:RG}
We now use RG to analyze the theory introduced in the previous section. Summing up Eqs.~(\ref{Spsi})-(\ref{Spsi-u}), we have

\begin{widetext}
\begin{align}
\mc S = \int d^4 x \left\{ \sum_{n=1}^N\bar\psi_n \left[ Z_\psi \g_0 \partial_0  + v_F \g_j \partial_j \right] \psi_n + {1\over 2}u_j\left[ \left(- Z_u^2\partial_0^2+\w_T^2 \right)\d_{jl} -c_T^2\left(\nabla^2\d_{jl}-\partial_j\partial_l  \right)-c_L^2\partial_j\partial_l \right]u_l \right. \nonumber \\ \left.
+ V \left(u_j^2 \right)^2 +{\ve_{\infty} \over 8\pi} \left( \partial_j \phi\right)^2 +ie \sum_{n=1}^N\bar\psi_n \g^0 \psi_n \phi+iQ\, \phi\, \partial_j u_j+\l \sum_{n=1}^N \bar \psi_{n} \g_0\g_j \psi_{n}\,u_j \right\}.\label{S_fixed}
\end{align}
\end{widetext}
The coefficients $Z_\psi$ and $Z_u$ account for the renormalization of the dynamical terms.

We apply the standard momentum-shell RG scheme~\cite{Fisher1974} by separating fields into short- and long-scale parts according to $\psi(\omega, \bs q) = \psi_>(\omega, \bs q) + \psi_<(\omega, \bs q)$ (analogously with fields $\bu$ and $\phi$), followed by the integrating out the high-energy part $\psi_>(\omega, \bs q)$ within an infinitesimal cylindrical momentum-frequency shell $\Lambda_0 e^{-\delta l} < q < \Lambda_0,$ $-\infty < \omega < \infty$. Here, $\Lambda_0$ is a momentum UV cutoff corresponding to the scale at which electron dispersion can be considered linear, and $l$ is ``RG time''. As the second step, we further rescale momenta, frequencies, and the long-wavelength parts of the fields according to

\begin{align}
&\bs q = e^{-\delta l} \bs q', \, \omega = e^{-z \delta l} \omega', \, \psi_<(\omega, \bs q) = e^{\eta_\psi \delta l} \psi'(\omega', \bs q'),  \nonumber \\ &\phi_<(\omega, \bs q) = e^{\eta_\phi \delta l} \phi'(\omega', \bs q'), \, \bu_<(\omega, \bs q) = e^{\eta_u \delta l} \bu'(\omega', \bs q'),
\end{align}
to restore the UV cutoff $e^{-\delta l} \Lambda_0$ back to $\Lambda_0$. Above, $z$ is the dynamical exponent, and $\eta_\psi$, $\eta_u$, $\eta_\phi$ are engineering field dimensions. This rescaling leads to the tree-level RG flows of the couplings after coarse-graining by the factor $e^{l}$ (the argument $l$ is suppressed for brevity):
 \begin{align}
&{ Z_{\psi}/ Z_\psi(0) } =  e^{(2\eta_\psi - 2z - 3)l  } ;\;\; { Z_{u}/ Z_u(0)} = e^{ (\eta_u -3z/2-3/2)l}
\nn \\
&{ c_{T,L}/ c_{T,L}(0)} =e^{(\eta_u-z/2-5/2)l}; \; {  v_F /v_F(0)} =e^{(2\eta_\psi - z - 4)l }
\nn\\
&{ \w_T/ \w_T(0)} =e^{(\eta_u -z/2-3/2)l} ;\;\;\;
{ V/ V(0)} =e^{(4 \eta_u -3z-9)l}\nn\\
&{ e /e(0)} =e^{(2 \eta_\psi+\eta_\phi -2z-6)l}\;\;\;;\;\;\;
{ Q/Q(0)} =e^{ (\eta_u +\eta_\phi -z-4)l}\nn\\
&{ \lambda/\l(0) } =e^{(2\eta_\psi+\eta_u -2z-6)l } \;\;; \;\;\; {\ve_\infty/\ve_\infty(0) = e^{(2\eta_\phi - 5 -z)l}}.
\end{align}
It should be mentioned that the choice of dynamical and field exponents is somewhat arbitrary here since it does not affect the flow of dimensionless coupling constants~\cite{Radzihovsky2011,Kozii2017}. The special choice $\w_T^2=0$, $Q=e=\l=0$, $\eta_\psi = 5/2$, $\eta_u = \eta_\phi = 3$, and $z=1$ makes the theory scale invariant. This particular choice is a noninteracting fixed point. Near this fixed point, $\w_T^2$ and $Q$ are relevant perturbations, while $e$ and $\lambda$ are marginal at the tree level. Since $\w_T^2$ is the tuning parameter for the FE transition, we will assume it to be small close to the critical point. In what follows, we focus on two distinct cases: The case of ionic crystals with $Q \ne 0$ and the case of covalent crystals with $Q = 0$.

\subsection{Ionic crystals ($Q \ne 0$) \label{sec:mu_ne_0}}
\subsubsection{Fixed point theory}
Near the noninteracting fixed point introduced above, $Q$ is relevant and, at the tree level, obeys the following RG equation:
\be
{d Q \over dl} = Q.
\ee
Thus, in the case of ionic crystals, $Q$ grows rapidly to strong coupling. Therefore, we should first derive the effective low-energy theory with large coupling $Q$ (of the order of UV cutoff) and then proceed to the RG analysis of the resulting theory. We can integrate out the Coulomb field $\phi$, which generates the following terms
\begin{align}
&\int D[\phi]e^{-\mc S_\phi - \mc S_{\psi \phi} - \mc S_{u\phi}} \propto e^{-\int_{\w,\bold{q}}\mc L'}\;\;;    \label{Eq:Frohlich} \\
& \mc L' ={2\pi \over\ve_{\infty} q^2}\left[ { e^2} |\d \rho_e(q)|^2 + {Q^2} |\bs q \cdot \bs u_q|^2 - {2 e Q i  } \bs q \cdot \bs u_q \,\d\rho_{e}(-q) \right]\nn
\end{align}
The first term is the standard Coulomb repulsion between electrons. The second term can be viewed as a phonon mass generated in the longitudinal sector (note that it is independent of the magnitude of the momentum). This mass generation is the well-known LO-TO splitting  in ionic crystals~\cite{mahan2013many}. Finally, the last term is the Fr\"ohlich coupling between the longitudinal phonon mode and electronic density.

{The generated mass term for the longitudinal phonon mode equals $\w_L \equiv \sqrt{4 \pi Q^2 / \ve_{\infty}  +\w_T^2 }$.  While the transverse mode becomes massless near the FE transition, $\omega_T\to 0$, we see that the longitudinal mode remains massive, since $Q \gg \omega_T$. Consequently, the LO mode can be further integrated out.  This procedure generates the standard dynamically screened Coulomb interaction between electrons
{
\be
\mc S_C = \frac12 \int {d \omega d^3 q \over (2\pi)^4} {4 \pi e^2 \over \ve(\w, q) q^2} |\d\rho_{e}|^2,\label{screened_Coulomb}
\ee}
where
\be
\ve(\w, q) = \ve_\infty {\w^2 + \w_L^2 +c_L^2 q^2  \over \w^2 + \w_T^2+c_L^2 q^2} \label{Eq:epsilonpolar}
\ee
is the dynamical dielectric constant, which manifestly satisfies the Lyddane-Sachs-Teller relation~\cite{AMbook,mahan2013many,lyddane1941polar}.

Close to the critical point, we have $\w_T \rightarrow 0$, which implies that the dielectric constant scales as $\ve(\w,q) \approx \ve_\infty \w_L^2/(\w^2+c_L^2 q^2)$ and diverges at low energies and momenta. (In the above estimate, we assumed that $Q$ already reached the RG scale at which $Q\sim \omega_L \sim c_L \Lambda_0$, where $\Lambda_0$ is the UV cutoff, implying that the $\omega^2 + c_L^2 q^2$ term can be neglected compared to $\omega_L^2$.) Thus, the effective Coulomb interaction between electrons becomes highly irrelevant and flows quickly to zero. It means that the Coulomb interaction is effectively screened by the longitudinal phonon mode. Eventually, the FE critical point is controlled by the following effective field theory
\begin{widetext}
\be
\mc S = \int d^4 x \bigg[\sum_{n=1}^N\bar\psi_n \left(  Z_\psi\g^0 \partial_0  + v_F \g^j \partial_j  \right) \psi_n +{1\over 2}u_j\left[- Z_u^2 \partial_0^2+\w_T^2 - c_T^2 \bs \nabla^2 \right]P_{jl}\,u_l +  V \left(u_j P_{jl} u_l \right)^2  + \l \sum_{n=1}^N P_{jl}\,u_l \bar \psi_{n} \g_0\g_j \psi_{n}\bigg]\label{eq:S_polar}
\ee
\end{widetext}
where
\be
P_{jl}(\bs q) = \d_{jl} - {q_j q_l /\bs  q^2}\label{projector}
\ee
is the projector to the plane transverse to $\bs q$. The couplings $V$ and $\l$ here are weakly renormalized after integrating out the longitudinal mode.

Before continuing, we would like to make a remark. We have presented a low-energy effective theory with the coupling between soft polar phonons and charged fermions given by Eq.~\eqref{Spsi-u}, which is not a long-range Coulomb interaction. This is a somewhat counter-intuitive result, as the phonon distortions generate huge dipolar moment that na\"{i}vely induces a long-ranged potential.
Consequently, one might expect that the Coulomb interaction between electrons and lattice distortions, arising from a deformation potential, is dominant over the direct coupling~\eqref{Spsi-u}. However, as we demonstrated above, the Coulomb forces that lead to the strong  Fr\"{o}hlich coupling between electrons and phonons (third term in Eq.~\eqref{Eq:Frohlich}) are also responsible for the generation of huge mass $\omega_L\propto Q$ for longitudinal phonons (second term in Eq.~\eqref{Eq:Frohlich}) and LO-TO splitting. As a result, the phonon mode that remains soft at the transition is precisely the transverse one, which does not generate a dipolar moment. This result is exact for an isotropic phonon dispersion. In the case of an anisotropic dispersion, there is always a finite mixture between the LO and TO modes leading to a remnant polarization in the soft phonon branch~\cite{Wolfle2018}. However, this remnant goes quickly to zero at small $q$ (as $q^2$) , rendering this coupling less relevant than Eq.~\eqref{Spsi-u}~\cite{ruhman-comment}.

\subsubsection{One-loop RG analysis}

Now we analyze the effective field theory for ionic crystals~(\ref{eq:S_polar}) within the one-loop RG approach. To get rid of the exponents $\eta_\psi,\, \eta_u,\, z$, which, in principle, can be chosen arbitrary, we focus on the dimensionless quantities which are independent of these engineering dimensions~\cite{Kozii2017,Radzihovsky2011}. First, we derive coupled RG equations for the ratio of the phonon to electron velocities $\zeta_T \equiv c_T Z_\psi / v_F Z_u$ and the dimensionless electron-phonon coupling constant  $\b \equiv \l^2/4 \pi c_T^2 v_F Z_\psi$ (the details of the calculation can be found in Appendices \ref{app:diags} and \ref{app:polar}):
\begin{align}\label{RG_eqs_polar}
&{d \zeta_T \over dl} =-{\zeta_T(1+\zeta_T)^2(1+\zeta_T^2)N - 8\zeta_T^2 \over 6\pi (1+\zeta_T)^2 }\b, \nonumber\\
&{d \b \over dl} ={(1+\zeta_T)^2 N-4 (1-\zeta_T)\zeta_T\over 3 \pi (1+ \zeta_T)^2}\b^2.
\end{align}

\begin{figure}[t!]
 \begin{center}
    \includegraphics[width=1\linewidth]{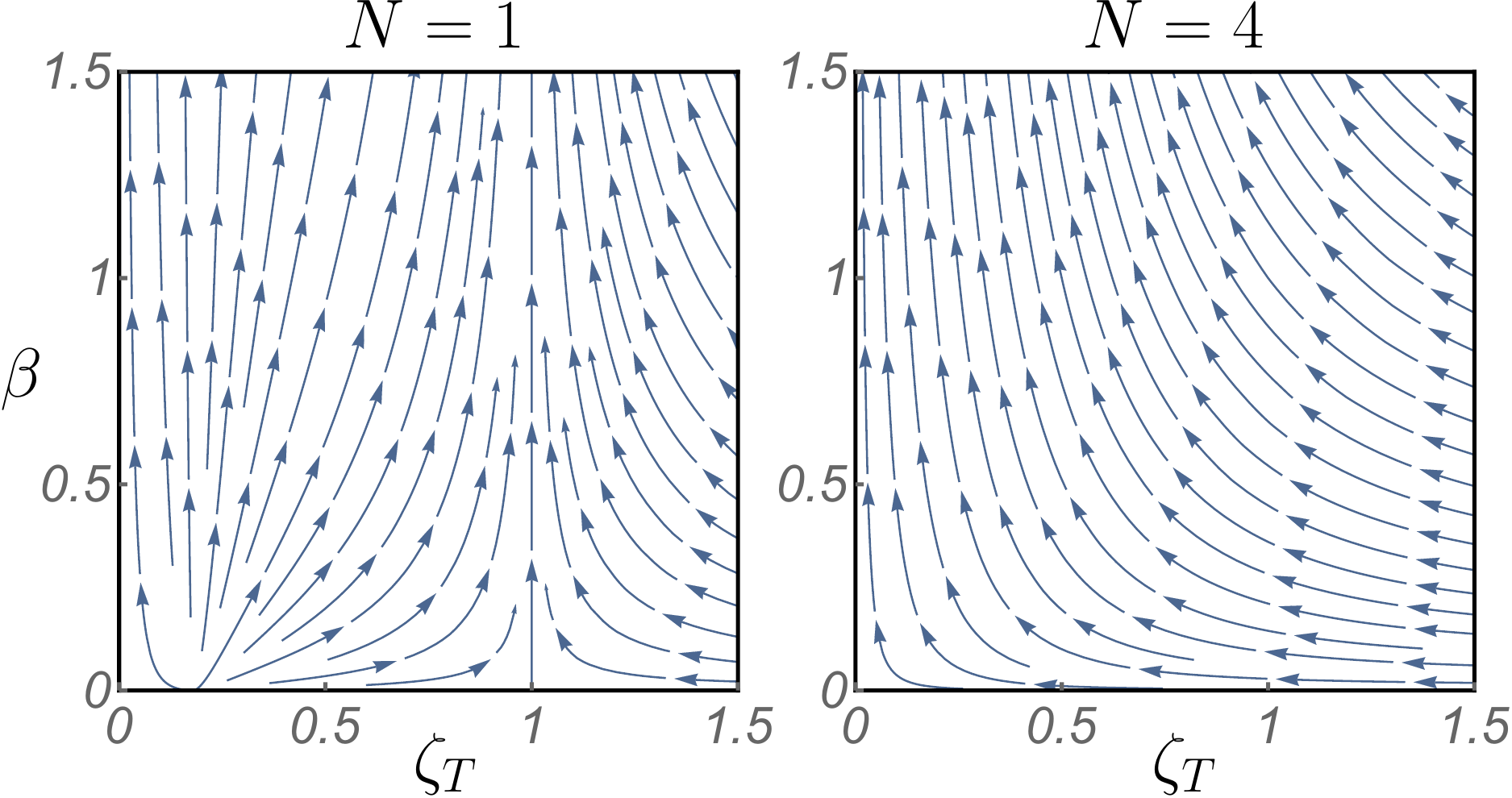}
 \end{center}
\caption{The RG flow  of the dimensionless electron-phonon coupling $\b$ and the velocity ratio $\zeta_T$ given by Eq.~\eqref{RG_eqs_polar} for a single Dirac cone (left) and for $N=4$ (right). In both cases, the flow is towards strong coupling, where the one-loop RG analysis breaks down. }
 \label{fig:flow}
\end{figure}

The most important result that can be extracted from these equations is that the electron-phonon coupling $\b$ flows to the strong-coupling regime, see Fig.~\ref{fig:flow}. Consequently, we conclude that the (3+1)D ferroelectric quantum critical point in a Dirac semimetal considered in this paper is generically a strongly-coupled problem, even if the original UV value of the coupling constant is small. This conclusion may be contrasted with standard QED in (3+1) dimensions, where the flow of the interaction is towards weak coupling, and the low-energy effective theory is the noninteracting Dirac fermion with renormalized parameters~\cite{isobe2012}. In the next section, we discuss the possible superconducting instabilities resulting from this flow to strong coupling.

Our RG equations were derived under the assumptions of the zero Dirac fermion mass and zero Fermi energy, while the one-loop approximation is valid provided the coupling remains small. Given the flow to the strong coupling, it is important to understand what stops the RG flows. Here we estimate the scale at which $\beta$ becomes of the order of 1,  and defer the discussion of a finite Dirac mass/Fermi energy to Sec.~\ref{sec:SC}. In realistic materials, the Fermi velocity is much bigger than the phonon velocity, thus, one can set $\zeta_T \approx 0$ in Eq.~(\ref{RG_eqs_polar}). Then, the equation for the flow of $\beta$ can be readily integrated. Completely neglecting the mass of the soft mode, $\omega_T \approx 0,$ we find that $\beta$ grows to $\sim O(1)$ at the RG scale $l_\beta = 3\pi/N\beta_0,$ which corresponds to the momentum scale
\be
\L_\b\sim \L_0 \exp\left(-\frac{3\pi}{N \beta_0}\right). \label{Eq:Lb}
\ee
Here $\beta_0 \ll 1$ is the initial UV value of the coupling constant at the scale $\Lambda_0$.

\begin{figure}
 \begin{center}
    \includegraphics[width=1\linewidth]{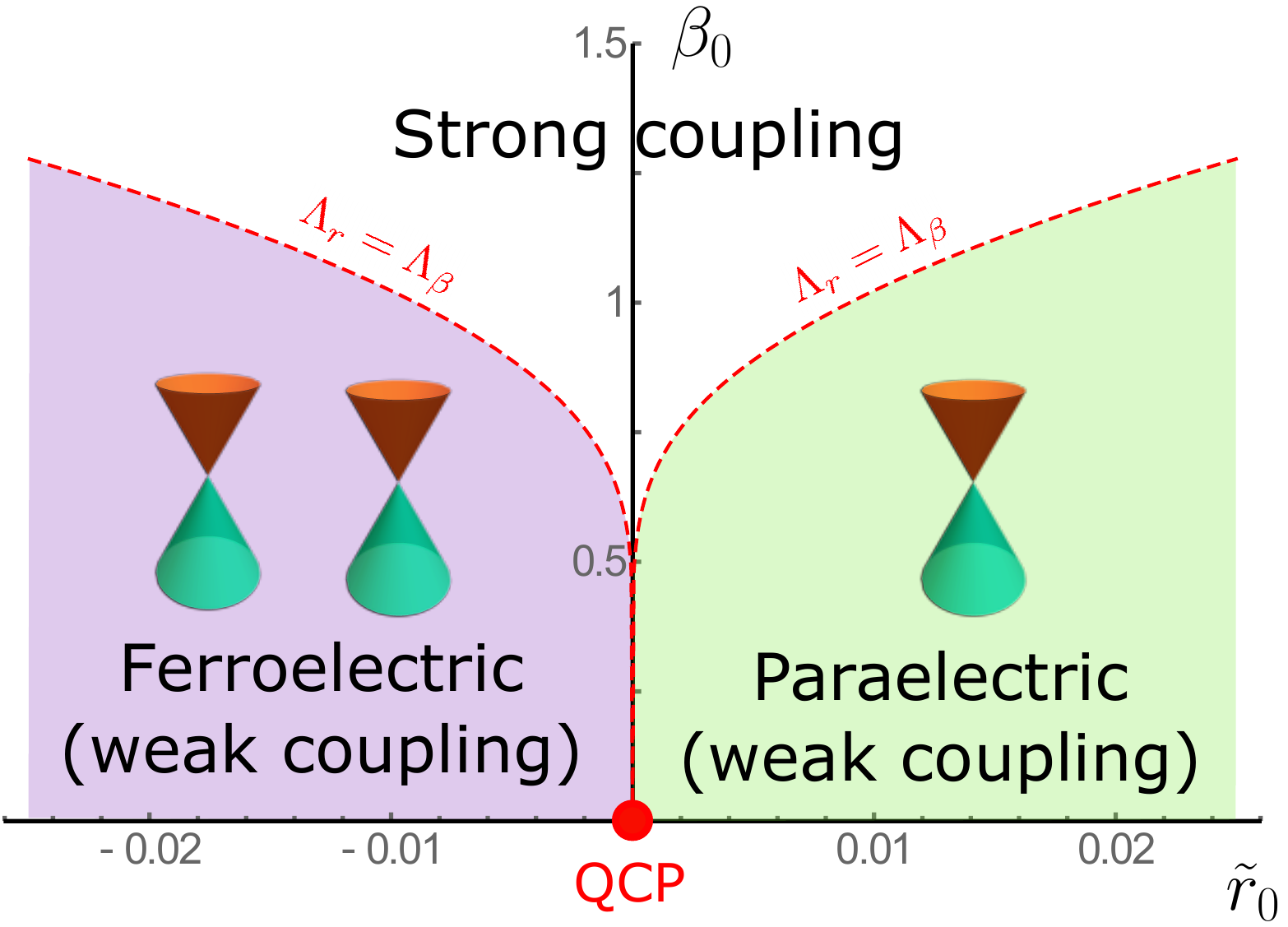}
 \end{center}
\caption{The phase diagram of a Dirac semimetal with $N=4$ close to a ferroelectric quantum critical point as a function of the bare values of the control parameter $\tilde r_0 = r_0 +(5\gamma_0/3\pi^2) - (2 N \beta_0/3\pi)$  and the electron-phonon coupling $\b_0$. The red dashed line separates the two regions, $\L_r >\L_\b$ and $\L_r <\L_\b$, corresponding to weak- and strong-coupling regimes, respectively. In the former region, the finite mass of phonons cuts off the RG flow before $\b$ reaches strong coupling, and the theory with renormalized parameters remains weak-coupled.  In the latter case, $\b$ flows to strong coupling before the system leaves the critical region. This regime is characterized by strong electron-phonon coupling and requires further study. The insets in each region schematically depict the dispersion close to the four $L$ points. Every Dirac cone in the paraelectric phase splits into two Weyl points in the ferroelectric phase.}
 \label{fig:phase_diag}
\end{figure}

Another natural scale that serves as a cutoff for our RG equations is set by the flow of the (dimensionless) mass of the transverse phonon mode $r \equiv \omega_T^2 / c_T^2 \Lambda_0^2$, which determines the critical region:

\be
\frac{d r}{dl} =  r \left( 2 + \frac{N \beta}{3\pi} \right) - \frac{4N \beta}{3\pi} + \frac{10 \gamma}{3\pi^2},
\ee
where $\gamma \equiv V/c_T^3 Z_u$ is the dimensionless phonon-phonon interaction. Assuming that $\beta$ and $\gamma$ are small compared to the UV value $r_0$, the solution of this equation with the exponential accuracy reads as $r \sim  r_0 e^{2l}$. The critical regime is determined by the condition $r \lesssim 1$, which corresponds to the RG scale $l_r \approx (1/2)\ln(1/|r_0|),$ or, equivalently, momentum scale \footnote{In case of finite (but small) $\beta_0$ and $\gamma_0$, $r_0$ in Eq.~(\ref{Eq:Lr}) should be replaced with $r_0 \to \tilde r_0= r_0 +(5\gamma_0/3\pi^2) - (2 N \beta_0/3\pi)$.}

\be
\L_r \sim \L_0 \exp\left[-\frac{1}{2}\ln(1/|r_0|)\right]=\L_0 \sqrt{|r_0|}.  \label{Eq:Lr}
\ee
If $\L_\b > \L_r$, the theory flows to the strong coupling regime before the phonon mode gets massive. Our RG equations are only applicable then down to $\L_\b$. In the opposite case, $\L_r > \L_\b$, the RG flow should be stopped at $\L_r$, where the transverse phonon mode becomes massive and can be integrated out. At this scale, the system leaves the critical regime, while the coupling between phonons and fermions still remains weak. The corresponding phase diagram is shown in Fig.~\ref{fig:phase_diag}. We will consider the latter case in more detail in the next section in context of superconductivity.

Finally, we discuss the flow of the dimensionless phonon-phonon interaction $\gamma \equiv V/c_T^3 Z_u$, which corresponds to the anharmonicity of the lattice oscillations:
\be
\frac{d \gamma}{dl} = \gamma \left[ \frac{N \beta (3 -\zeta_T^2)}{6 \pi} - \frac{17 \gamma}{5 \pi^2}  \right] - \frac{2 N \beta^2 \zeta_T}{3}. \label{Eq:gammaflow}
\ee
This equation, again, can be easily analyzed in the physical case $\zeta_T \approx 0$. Then, since $\beta$ is a marginally relevant parameter, $\gamma$ eventually also flows to strong coupling. It is straightforward to show, however, that this flow does not introduce any new cutoff, as $\gamma$ can reach order 1 no sooner than at $\Lambda_\beta$ given by Eq.~(\ref{Eq:Lb}), see Fig.~\ref{fig:gammaflow}. This scenario is realized in the large-$N$ limit, i.e., when the term proportional to $\propto \gamma^2$ on the right-hand side of Eq.~(\ref{Eq:gammaflow}) can be neglected. It is also interesting to note that sufficiently large $\zeta_T$ in Eq.~(\ref{Eq:gammaflow}) can, in principle, drive $\gamma$ negative, thus indicating a first-order transition into the ferroelectric state. Since we consider $\zeta_T \sim 1$ hardly realizable in real physical systems, we do not study this possibility in detail here.

\begin{figure}
 \begin{center}
    \includegraphics[width=1\linewidth]{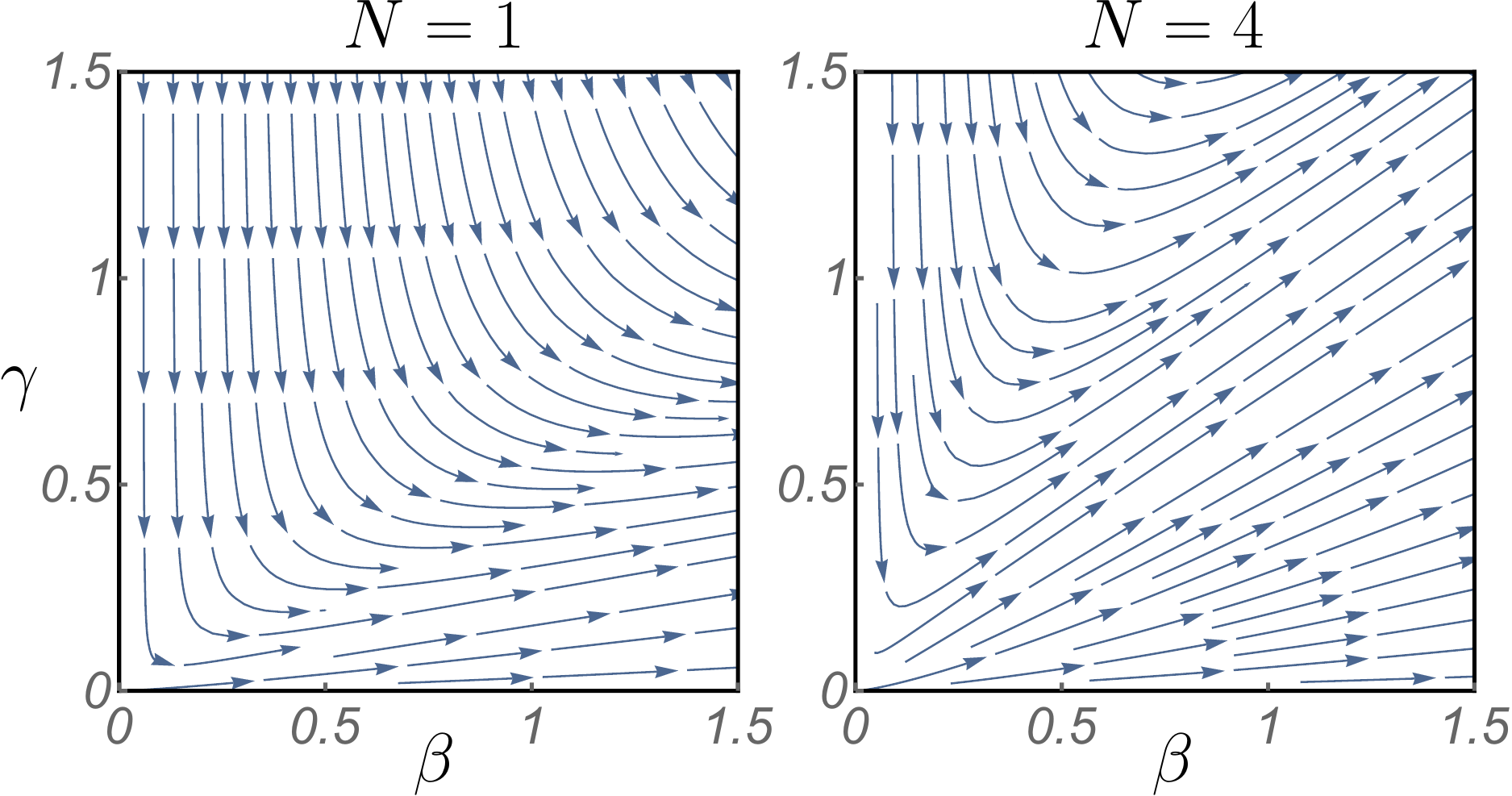}
 \end{center}
\caption{The RG flow of the dimensionless phonon-phonon coupling $\g$ and the dimensionless electron-phonon coupling $\b$ corresponding to Eqs.~(\ref{RG_eqs_polar}) and~(\ref{Eq:gammaflow}) in the limit $\zeta_T = 0$. The left panel represents the case of a single fermionic flavor $N=1$, and the right panel is for $N=4$. The scale when $\gamma$ reaches $\sim1$ never exceeds $\Lambda_\beta$. }
 \label{fig:gammaflow}
\end{figure}

Another interesting result that can be inferred from the RG equations is the flow of the electron and phonon velocities (here we fix the dynamical critical exponent $z = 1$):

 \begin{align}
&{d(v_F/Z_\psi) \over dl} = -{ 4\zeta_T\b \over 3 \pi (1+\zeta_T)^2}\,\frac{v_F}{Z_\psi}, \nonumber \\
&{d(c_T/Z_u) \over dl} = -{(1+ \zeta_T^2)\beta N  \over 6 \pi}\,\frac{c_T}{Z_u}.
\end{align}
We see that one of the physical properties of the ferroelectric critical point in Dirac materials is the reduction of the velocities under RG for both the transverse phonon modes and the Dirac fermions. Furthermore, as is shown in Fig.~\ref{fig:flow}, for $N=1$, the velocity ratio $\zeta_T$ flows to one of two possible values $\zeta_0 = 0$ or $\zeta_1 = 1$, depending on whether the initial value of $\zeta_T$ is smaller or larger than $\zeta_m =  t / 3^{2/3} -2/(3^{1/3}t)-1 \approx 0.18$, respectively, with $t = (18+2\sqrt{87})^{1/3}$. If $N>1$, the flow is always towards $\zeta_0=0$.

So far we only considered a rotationally symmetric model with isotropic electron and phonon velocities. For $N>1$, however, there is no symmetry that forbids anisotropic terms which manifest the symmetry of the underlying lattice. Nevertheless, the accounting for these terms does not modify main qualitative results described above. Hence, we focus on the isotropic case for the rest of the paper for simplicity, and defer the discussion of  possible anisotropies to Appendix~\ref{App:Anisotropy}.

\subsection{Covalent crystals ($Q = 0$) \label{Sec:CovalentRG}}

Now we perform similar RG analysis for covalent crystals, exemplified by elemental bismuth. The ``ferroelectric'' phase in these materials is characterized by broken inversion symmetry but not by a spontaneous dipolar moment of the lattice, because the optical phonon distortion $\bs u$ generates a negligible amount of polarization in covalent crystals.

While the main qualitative results, such as the flow to strong coupling, in this case are the same as for ionic crystals, certain important differences should be discussed. In particular, the absence of lattice polarization implies that the effective theory for covalent crystals is given by Eq.~\eqref{S_fixed} with $Q=0$. As a result of this important difference, the argumentation of Section~\ref{sec:mu_ne_0} about the screening of Coulomb interaction by massive longitudinal phonons no longer holds. Instead, one should keep track of the flows of the parameters $\ve_\infty$ and $e$, in addition to those considered in Eq.~\eqref{RG_eqs_polar}. Focusing again on dimensionless parameters that do not depend on engineering dimensions $\eta_\psi,\, \eta_u, \, \eta_\phi,$ and $z$, we find the following set of coupled one-loop RG equations:

\begin{align}
{d \b \over dl} = & {N\b^2 \over 3 \pi} +\frac{4\beta^2 \zeta_T(\zeta_T-1)}{3\pi (1+\zeta_T)^2} - \frac{2\beta^2 \zeta_T^2}{\pi\zeta_L(1+\zeta_L)^2}, \nn \\
{d \a \over dl} =&-{2 (N+1)\a^2 \over 3 \pi}+ \frac{4\a \beta \zeta_T}{3\pi (1+\zeta_T)^2}+\frac{2\a\beta \zeta_T^2 (3+\zeta_L)}{3 \pi \zeta_L (1+\zeta_L)^2},\nn \\
{d \zeta_T \over dl} = & -\frac{N\b(1+\zeta_T^2)\zeta_T}{6\pi}+\frac{4\b\zeta_T^2}{3\pi (1+\zeta_T)^2 }\nn\\& + \frac{2\beta \zeta_T^3 (3 + \zeta_L)}{3 \pi \zeta_L (1+\zeta_L)^2}-\frac{2\alpha\zeta_T}{3\pi},\nn\\
{d \zeta_L \over dl} = &{N\b \zeta_T^2(1-\zeta_L^2)\over 6\pi \zeta_L} + \frac{4\beta \zeta_T \zeta_L}{3\pi  (1+\zeta_T)^2}\nn \\ &+ \frac{2\beta \zeta_T^2 (3 + \zeta_L)}{3 \pi (1+\zeta_L)^2}-\frac{2\alpha\zeta_L}{3\pi}, \label{Eq:S_nonpolar}
\end{align}
where we defined $\b \equiv \l^2  / 4\pi v_F c_T^2 Z_\psi$, $\a \equiv e^2 / \ve_{\infty} v_F Z_\psi$,  $\zeta_T \equiv c_T Z_\psi / v_F Z_u$, and $\zeta_L \equiv c_L Z_\psi / v_F Z_u$.

Since both the longitudinal and the transverse phonon modes become massless at the transition in covalent crystals, they should be treated on equal footing. Consequently, one could in principle consider two (not independent) dimensionless couplings $\b=\l^2 / 4\pi v_F c_T^2 Z_\psi $ and $\tilde\b= \l^2 / 4\pi v_F c_L^2 Z_\psi $, which quantify the electron-electron interaction strength mediated by the transverse phonons and the longitudinal phonons, respectively. It is straightforward to show, however, that, in the physical limit $\zeta_T \sim \zeta_L \ll 1$, $\tilde\b$ is marginally irrelevant, while $\beta$ flows to strong coupling. Indeed, in this limit, first two equations of Eq.~(\ref{Eq:S_nonpolar}) take form (assuming also that $\beta \zeta_T \ll \alpha$)

\begin{align}\label{RG_eqs_non_polar2}
&{d \b\over dl} ={N\b^2\over 3\pi}, \nn \\
&{d \a \over dl} =-{2 (N+1)\a^2 \over 3 \pi},
\end{align}
while analogous equation for $\tilde \b$ would read as
\be
\frac{d\tilde\b}{dl} = - \frac{N\tilde \b^2}{3\pi}. \label{Eq:betatilde}
\ee
We note that in this limit, the renormalization of the fine structure constant $\a$ is identical to that in standard QED in (3+1) dimensions~\cite{Hosur,isobe2012}. Similarly to the case of ionic crystals, electron-phonon coupling $\beta$ flows to the strong-coupling regime, while the Coulomb interaction $\alpha$ becomes suppressed under RG. The important difference, however, is that now $\alpha$ is only marginally irrelevant and flows to zero much slower. The reason for this difference is that the Coulomb screening in covalent crystals is due to interband (between particle and hole bands) transitions, which is much weaker than the screening by the lattice polarization in ionic crystals.

The flow of the phonon velocities can also be easily studied in the limit $\zeta_T \sim \zeta_L \ll 1$. Analogously to ionic crystals, $\zeta_T$ flows to zero in this regime. The flow of $\zeta_L$, on the other hand, is sensitive to the number of flavors $N$ and to the initial conditions, as well as to the scale that stops RG. For instance, at sufficiently large $N$, $\zeta_L$ is increased under RG.

Finally, the flows of the phonon-phonon coupling $\gamma$ and the phonon mass $r$ are qualitatively similar to the case of ionic crystals, so we do not consider them in detail here.

\section{Superconductivity \label{sec:SC} }

In the previous section, we have analyzed the RG flow of the electron-phonon coupling near a ferroelectric quantum critical point. We found that, generically, the critical point is unstable and flows to strong electron-phonon coupling, while the Coulomb interaction flows to weak coupling. As a result, we anticipate that the effective electron-electron attraction mediated by the ferroelectric phonon modes will become dominant over the Coulomb repulsion. Hence, the natural next step in our work is to apply this result to study superconductivity.

We emphasize that in our scenario, both for ionic and covalent crystals, the enhancement of the attractive interaction over the Coulomb repulsion does not require finite electron density, in contrast to the Anderson-Morel theory. Nonetheless, this does not imply that the superconducting transition temperature does not depend on density. At least at weak coupling, $\b \lesssim 1$,  a finite density of states is essential to obtain a finite $T_c$. Therefore, we will now relax our previous assumption about the Fermi energy exactly at the Dirac point and assume a nonzero Fermi momentum $k_F$. As before, we separately consider the cases of ionic and covalent crystals. We also focus on the paraelectric side of the transition, i.e., consider systems possessing both time-reversal and inversion symmetries in the normal state.

\subsection{Ionic crystals}

As we have shown in Sec.~\ref{sec:RG}, one can define two scales $\L_\b$ and $\L_r$ given by Eqs.~\eqref{Eq:Lb} and~\eqref{Eq:Lr}, which correspond to the divergence of the electron-phonon coupling $\b$ and the phonon mass $r$, respectively. When $\L_r > \L_\b$, $r$ diverges first, and the flow is terminated before $\b$ reaches strong coupling (this regime is denoted by the shaded regions in Fig.~\ref{fig:phase_diag}). In what follows, we consider this weak-coupling limit, where the BCS approach is applicable, and leave the strong-coupling scenario $\L_\b > \L_r$ for a future work.

The additional scale we have introduced, $k_F$, can, in principle, also put the flow to a halt when the running scale $\L(l) = \L_0\exp(-l)$ becomes of the order of $k_F$. Thus, depending on the ratio between $k_F$ and $\L_r$, one may again consider two cases. The first case, $k_F \gtrsim \L_r$, is close to the standard Anderson-Morel scenario with the phonon-associated scale $(\omega_T/c_T)\vert_{l=0}$ being smaller than $k_F$, and we do not consider it here in detail. Since we are interested in  understanding  superconductivity at very low density, we focus on the opposite limit $\L_r \gtrsim k_F$. In this case, the screening of the Coulomb repulsion by longitudinal phonons occurs well above the Fermi scale, as discussed below Eq.~\eqref{Eq:epsilonpolar}, and we obtain a Fermi liquid with {\it static} phonon-mediated attraction.  The inequality $\L_r \gtrsim k_F$ also implies that (at finite density) the system is away from the immediate vicinity of the critical point; the behavior exactly at criticality will be considered in a separate work.

To obtain an effective low-energy interaction, we allow the system to flow according to the RG equations derived in Sec.~\ref{sec:mu_ne_0} until it reaches the scale $\L_r$. We then use Eq.~\eqref{eq:S_polar} to integrate out the transverse phonon mode, which is massive at this scale, with the effective propagator that can be considered frequency- and momentum-independent. This procedure results in the attractive interaction Hamiltonian

 \be
\mc H_{FE} = -{ 2\pi v_F^* \b^*\over \L_0^2}\sum_{k,k',q} P_{jl}(\bs q)\left(\psi^\dag_{k+q} \g_j \psi_k\right)\left(\psi^\dag_{k'-q} \g_l \psi_{k'}\right), \label{FE_interaction}
\ee
where the effective interaction constant

 \be
 \b^{*} \equiv \b(l_r) = {\b_0 \over 1-{\b_0 N\over 3\pi} \log{\L_0\over \L_r}} \label{Eq:beta*}
 \ee
is obtained from Eq.~\eqref{RG_eqs_polar} in the limit of $\zeta_T \ll 1$,  and we also defined the renormalized Fermi velocity $v_F^* \equiv v_F(l_r).$ To make the analysis similar to the conventional BCS at this point, we write
Eq.~\eqref{FE_interaction} in the Hamiltonian formalism (and use $\psi^\dagger$ instead of $\bar \psi$). Such rewriting is possible since, at the scale $\L_r$, the phonon-mediated interaction can be considered static, $\omega \lesssim \omega_T$, analogously to BCS theory.

\subsubsection{Projection onto the Fermi level}

Now we analyze the superconducting instabilities due to interaction~\eqref{FE_interaction}. We assume that the Fermi energy $\ve_F = v_F k_F$ is much larger than the superconducting gap, $\ve_F \gg \Delta$, hence, the conventional weak-coupling BCS-like treatment is applicable. In this case, it is convenient to project all operators onto the band where the Fermi level resides, thus significantly simplifying the model by reducing it from the original four-orbital to effective two-orbital.  In the paraelectric phase, the only case we consider in this Section, both time-reversal and inversion symmetry are present in the normal state, hence, all energy bands remain double degenerate even in presence of strong spin-orbit coupling. The electron states are characterized by a two-component spinor $c_{\bs k} = \left[ c_1(\bs k), c_2(\bs k) \right]^T$. In the presence of spin-orbit coupling, however, components $c_{1,2}$ are not spin eigenstates anymore, but rather eigenstates in some band basis. The choice of this basis is not unique. For concreteness, we choose the so-called manifestly covariant Bloch basis (MCBB), in which  $\left[ c_1(\bs k), c_2(\bs k) \right]^T$ transforms as an ordinary $SU(2)$ spin-1/2~\cite{Fu2015}. To find this basis, we diagonalize Hamiltonian which corresponds to Eq.~\eqref{Spsi}, and choose the band eigenstates to be fully spin-polarized along the $z$-axis at the origin of the point group symmetry operations (see also Refs.~\cite{Kozii2015} and~\cite{Venderbos2017} for more details). The eigenvectors $b_1(\bs k)$ and $b_2(\bs k)$ in the MCBB that correspond to the states near Fermi energy are given by
\be
b_1(\bs k) = \frac12 \left( \begin{array}{c} \eta - \hat k_z \\ -  \hat k_+ \\ \eta + \hat k_z \\ \hat k_+    \end{array}  \right), \qquad b_2(\bs k) = \frac12 \left( \begin{array}{c} -\hat k_- \\ \eta + \hat k_z \\ \hat k_- \\ \eta - \hat k_z  \end{array}  \right), \label{Eq:MCBB}
\ee
where $\eta = \pm 1$ corresponds to the conduction/valence bands, respectively, and we defined $\hat k_\pm = (k_x \pm i k_y)/k$. The mapping onto the MCBB then simply implies the transformation $\psi(\bs k) \to b_1(\bs k) c_1(\bs k) + b_2(\bs k) c_2(\bs k)$, and can schematically be written as $\psi(\bs k) = \mc Q_\eta(\bs k) c(\bs k)$, where $\mc Q_\eta(\bs k)$ is a projector onto MCBB. It is straightforward to show then that the Dirac bilinear $ \g_j $, which couples to a soft phonon mode, projects onto
\be
M_{\bs p, \bs k}^j=\mc Q_\eta^\dag(\bs p) \g_j  \mc Q_\eta(\bs k) = {\eta\over 2} \left[i(\hat p_j - \hat k_j) + (\hat k_l+\hat p_l)\s_m \e_{lmj}  \right], \label{bilinear}
\ee
where $\e_{lmj}$ is the Levi-Civita tensor, $\sigma_j$ here are Pauli matrices acting in the MCBB, and  we defined $\hat k_j \equiv k_j /k.$

The effective interaction~\eqref{FE_interaction} projected onto the  band with the Fermi level has form (here we use the notation $\widehat{\bs k} \equiv {\bs k}/k$)

\begin{widetext}
\begin{align}
\mc H_{FE}^T \simeq  -  {\pi v_F^* \b^* \over  2\L_0^2}  \sum_{\bs k,\bs k',\bs q}P_{jl}(\bs q)
\bigg\{c_{\bs k + \bs q}^\dag&\left[i\left( \widehat{\bs k+ \bs q} - \widehat{\bs k}   \right) + \left(\widehat{\bs k } + \widehat{\bs k +\bs q} \right)\times \bs \s\right]_j c_{\bs k}\bigg\} \times \nn \\ &\left\{c_{\bs k'- \bs q}^\dag\left[i\left( \widehat{\bs k' - \bs q} - \widehat{\bs k'}   \right) + \left({\widehat{\bs k '}} + {\widehat{\bs k' -\bs q}}\right)\times \bs \s\right]_lc_{\bs k'}\right\}. \label{HFE}
\end{align}
\end{widetext}


\subsubsection{Pairing channels and transition temperature}
To demonstrate the superconducting instabilities, we now decompose interaction~\eqref{HFE} into pairing channels, analogously to how it has been done in Ref.~\cite{Kozii2015}. The time-reversal invariant superconducting order parameter generally takes form

\be
\hat F^\dagger = \sum_{\bs k, \alpha \beta \gamma} \epsilon_{\beta \gamma} F_{\alpha \beta}(\bs k) c^\dagger_{\bs k \alpha} c^\dagger_{-\bs k \gamma},
\ee
where, again, $\epsilon_{\beta\gamma}$ is the Levi-Civita symbol. In systems with strong spin-orbit coupling, spin $S$ and orbital angular momentum $L$ are not good quantum numbers. Instead, in systems with $O(3)$ symmetry considered here, all possible orders are characterized by the total angular momentum $J = L + S$. As was shown in Refs.~\cite{Fu2015} and~\cite{Kozii2015}, the form-factors $F_{\alpha \beta}(\bs k)$ up to order $J=1$ have the form shown in Table~\ref{tab:SC_decomp}. $L=0$ state $F_0$ corresponds to the conventional $s$-wave pairing with $J=0$, while $L=1$ sates are odd-parity $p$-wave, and transform as a pseudoscalar ($F_1$ with $J=0$) and a vector ($F_2^j$ with $J=1$) under the symmetry operations.

\begin {table}[t!]
\caption { The decomposition of the phonon-mediated interactions \eqref{HFE} and \eqref{HFEL} into the BCS channels $F_n$ with total angular momentum $J=0$ ($a_0$ and $a_1$) and $J=1$ ($a_2$), see Eqs.~\eqref{Eq:HBCS} and~\eqref{Eq:HLdec}.  Positive coefficients $a_i$ imply the attraction in the corresponding BCS channels.} \label{tab:SC_decomp}
    \begin{tabular}{| l | c |  c |c |c |}
    \hline
    Form-Factor  &  $\mc P$                                                                               & Coefficient  & Transverse  & Longitudinal    \\ \hline
     $F_0(\bs k) = I $  &  even                                                          &$a_0$&   1  & 2  \\ \hline
     $F_1(\bs k) =  \hat{ \bs k}\cdot \bs \s  $  &  odd                  &$a_1$&  $-1$ & $-2$  \\ \hline
     $F_2^{j}(\bs k) = \left(\hat{\bs k}\times \bs \s\right)^j $ & odd                 &$a_2$&  1/2  & $-3/2$  \\ \hline
    \end{tabular}
\end {table}

Next, we restrict the effective interaction~\eqref{HFE} to the Cooper channel with the zero total momentum by keeping terms with $\bs k' = -\bs k$ only. Focusing on the states near the Fermi surface, $|\bs k| \approx |\bs k'| \approx |\bs k + \bs q| \approx |\bs k' - \bs q| \approx k_F$, it is straightforward to decompose Eq.~\eqref{HFE} into the pairing channels $F_n$~\cite{Kozii2015}:
\be
\mc H_{FE}^T \approx  -{\pi v_F^* \b^* \over  2\L_0^2}\sum_{n=0}^{2}a_n^T \sum_{j} \hat F_{n}^{j\dagger} \hat F_n^{j}+\ldots, \label{Eq:HBCS}
\ee
where coefficients $a_n^T$ are listed in Table~\ref{tab:SC_decomp}. The ellipsis on the right-hand side of Eq.~\eqref{Eq:HBCS} denotes terms with $J>1$~\footnote{The pairing channels in the odd-parity sector (which always implies $S=1$) with $J>1$ generally have contributions from the terms with  $L = J-S$ and $L = J+S$ orbital angular momenta. As a result, the decomposition into these channels is interaction-dependent.}. The contribution from these terms is numerically small, and we do not consider it in this paper.

Up to order $J=1,$ only two channels are attractive and lead to a superconducting instability: the scalar $\hat F_0$ with $a_0^T = 1$ and the vector $\hat F_2^j$ with $a^T_2 = 1/2$.
We thus conclude that pairing in the $s$-wave channel is the most dominant superconducting instability.

The transition temperature $T_{c}$ (for a given $n$)  can be estimated from Eq.~\eqref{Eq:HBCS} using the usual gap equation~\cite{MineevSamokhin}:
{
\be
\delta_{ij} = \frac{\pi v_F^* \beta^*}{ 2\Lambda_0^2}a^T_n \sum_{\bs k} \tr [F^i_n(\bs k) F^j_n(\bs k)] \frac{\tanh(\xi_{\bs k}/2T_{c})}{\xi_{\bs k}}.  \label{Eq:gapequation}
\ee
In case of the most attractive $s$-wave channel, it takes form
\be
1 \approx \frac{ 2\pi v_F^* \beta^* a^T_0 \nu^*}{\Lambda_0^2} \int_{\sim T_{c}}^{\sim \ve_F} \frac{d \xi}\xi,
\ee
where $\nu^* = k_F^2(l_r)/2\pi^2 v_F(l_r)$ is the density of states at the Fermi energy per one spin projection per one Dirac node, with all quantities entering it  taken at the RG scale $l_r$. We emphasize that the upper cutoff in Eq.~\eqref{Eq:gapequation} is not the phonon frequency, as in the standard BCS theory, but given by the Fermi energy. This situation is somewhat analogous to the superfluidity in a charge-neutral Fermi liquid, studied in Ref.~\cite{GorkovMelikGarkhudarov}. We estimate transition temperature from Eq.~\eqref{Eq:gapequation} as

\begin{align}
&T_{c} \sim \ve_F \exp\left(  - \frac{\Lambda_0^2}{ 2\pi v_F^* \beta^* a_0^T \nu^*}  \right) = \nonumber \\ &= \ve_F \exp \left( - \frac{\pi \Lambda_r^2}{k_F^2 \beta^*} \right)     =\ve_F \exp \left( - \frac{\pi \omega_{T0}^2}{ k_F^2 c_{T0}^2 \beta^*}  \right). \label{Eq:Tc0}
\end{align}
The parameters $k_F$, $c_{T0}$, and $\omega_{T0}$ in this equation are the original (UV) values of the Fermi momentum, phonon velocity, and phonon mass, respectively, while $\beta_*$ is renormalized according to Eq.~\eqref{Eq:beta*},  and we used $k_F(l) = k_F e^l$.  In particular, $k_F$ is related to the total electron density $n_0$ and the number of Dirac nodes $N$ as $k_F=(3\pi^2n_0/N)^{1/3}$. We see that the proximity to the ferroelectric critical point leads to a significant enhancement of $T_c$. To emphasize  this point, we rewrite Eq.~\eqref{Eq:Tc0} in the form

\be
T_{c} \sim T_{c0} \left(\frac1{r_0}  \right)^\delta \gg T_{c0}, \qquad \delta = \frac{N \omega_{T0}^2}{6 k_F^2 c_{T0}^2} \gg 1, \label{Eq:Tcenhanced}
\ee
where $T_{c0} \sim \ve_F \exp\left( - \pi \omega_{T0}^2 /  k_F^2 c_{T0}^2 \beta_0 \right)$ is the estimate for a transition temperature that we would obtain without taking into account the critical nature of the ferroelectric fluctuations. We see that, even within the weak-coupling approximation $\beta_0 \ln (\Lambda_0/\Lambda_r) \lesssim 1$,  we obtain huge enhancement of the transition temperature by a factor of  $(1/r_0)^\delta$ due to the renormalization of the coupling $\beta$. This result is to some extent similar to the enhancement of $T_c$ by the critical nematic fluctuations obtained in Ref.~\cite{Lederer2015}.

We see from Eq.~(\ref{Eq:Tc0}) that the transition temperature is exponentially sensitive to $\beta^*$ and $k_F/\Lambda_r$, both of which can be made of the order of 1 close to the FE QCP. We estimate the  magnitude of $T_c$ given by Eq.~\eqref{Eq:Tc0} using parameters of a realistic system, such as Pb$_{1-x}$Sn$_x$Te. We assume that the system is close enough to the QCP such that $\b^*$ gets sufficiently renormalized and becomes of the order of one (in particular, we take $\b^* = 1$, which is controlled by the logarithmic divergence in Eq.~\eqref{Eq:beta*}). The phonon velocity can be estimated as $c_{T0} = 3\times 10^3$~m/s~\cite{jacobsen2013sound,an2008ab}, and we take a small (since we are close to the critical point)  phonon mass  $\w_{T0} = 0.35$~meV. The low-energy electronic structure of Pb$_{1-x}$Sn$_x$Te is given by $N=4$ Dirac cones with a typical Fermi velocity $v_F \simeq 10^6$~m/s~\cite{assaf2016massive}. For the electron density $n_0 = 2\times 10^{17}$~cm$^{-3}$, we find from Eq.~(\ref{Eq:Tc0}) $T_c \approx 440$~mK with $\ve_F\approx870$~K and $k_F / \L_r \approx 0.64$. Analogously, for $n_0 = 4\times10^{17}$~cm$^{-3}$, we obtain $T_c \approx 9.1$~K with $\ve_F\approx 1100$~K and $k_F / \L_r \approx 0.81$. Despite the fact that the values of $T_c$ obtained above are only very rough estimates, we conclude that Eq.~\eqref{Eq:Tc0} may lead to a significant transition temperature even for a very low density of electrons. We emphasize again that this result is possible because the critical phonon modes  can couple to electrons at zero momentum transfer (as opposed to the gradient coupling), see Eq.~(\ref{Spsi-u}), which allows for the range of the TO phonon-mediated attraction to become comparable to the distance between electrons. Indeed, in this case, the phonon-mediated interaction range is $\Lambda_r^{-1}$, which, sufficiently close to the critical point, becomes of the same order as the interparticle distance $k_F^{-1}$. This is in perfect agreement with the result by Gurevich et al.~\cite{Gurevich1962}, who pointed out that low-density superconductivity necessarily requires a sufficiently long-ranged attractive interaction.

Finally, we estimate the temperature that would correspond to a transition into the $p$-wave superconducting state $T_{cp}$:

\be
T_{cp} \sim \ve_F \exp\left(  - \frac{3\Lambda_0^2}{ 4 \pi v_F^* \beta^* a^T_2 \nu^*}  \right) = \ve_F \exp \left( - \frac{3\pi \omega_{T0}^2}{k_F^2 c_{T0}^2 \beta^*}  \right). \label{Eq:Tcp}
\ee
An additional factor of $3/2$ in the exponent appears due to the averaging over the directions of vector $\bs k$ in Eq.~\eqref{Eq:gapequation}. $T_{cp}$ is exponentially smaller than $T_c$, and, consequently, $p$-wave superconducting phase seems unreachable within the present scenario. However, we demonstrate in the next section that the presence of the repulsive Coulomb interaction can, under certain conditions, suppress $s$-wave channel and drive a system into the odd-parity $p$-wave superconducting state.

\subsection{Covalent crystals}
Our analysis of superconductivity in covalent crystals is similar to the ionic case. There are, however, two important differences. First, the longitudinal phonon mode also becomes soft at a ferroelectric transition, and, consequently, there will be an additional contribution to the effective electron-electron interaction mediated by a longitudinal mode. Second, the Coulomb repulsion is not screened by the lattice polarization, but only by the interband transitions. Consequently, as we showed in Sec.~\ref{Sec:CovalentRG}, the correspondent coupling constants $\tilde \beta$ and $\alpha$ are marginally irrelevant, see Eqs.~\eqref{RG_eqs_non_polar2} and~\eqref{Eq:betatilde}. They flow to zero only logarithmically upon RG and, thus, should be taken into account in the weak-coupling regime we are considering here. As we show below in Sec.~\ref{Sec:p-wave}, the inclusion of the Coulomb interaction allows one, upon proper tuning of the coupling constants, to realize a $p$-wave superconductor.

The effective electron-electron interaction due to longitudinal phonons projected onto the Fermi level  has the same form as Eq.~\eqref{FE_interaction}, but with the substitution $P_{ij}(\bs q) \to \delta_{ij} - P_{ij}(\bs q)$
and $\beta^* \to \tilde \beta^*,$
\begin{widetext}
\begin{align}
\mc H_{FE}^L \simeq  -  {\pi v_F^* \tilde\b^* \over  2 \L_0^2}  \sum_{\bs k,\bs k',\bs q}\left[\delta_{jl}-P_{jl}(\bs q)\right]
&\left\{c_{\bs k+ \bs q}^\dag\left[i\left(\widehat{\bs k +\bs q}- \widehat{\bs k } \right)+\left(\widehat{\bs k } + \widehat{\bs k +\bs q} \right)\times \bs \s\right]_j c_{\bs k}\right\} \times \nonumber \\ &\left\{c_{\bs k'- \bs q}^\dag\left[i \left(\widehat{\bs k' -\bs q} - \widehat{\bs k' }\right)+\left({\widehat{\bs k '}} + {\widehat{\bs k' -\bs q}}\right)\times \bs \s\right]_lc_{\bs k'}\right\}\label{HFEL}
\end{align}
\end{widetext}
where $\tilde \beta^*$ is given by (see Eq.~\eqref{Eq:betatilde})
\be
\tilde \b^{*} \equiv \tilde \b(l_r) = {\b_0 \over 1+{\b_0 N\over 3\pi} \log{\L_0\over \L_r}} \label{Eq:tildebeta*},
 \ee
and, again, we used the notation $\widehat{\bs k} \equiv {\bs k}/k$. The decomposition into the pairing channels has form similar to Eq.~\eqref{Eq:HBCS}, with $\beta^*$ substituted by $\tilde \beta^*$

\be
\mc H_{FE}^L\approx  -{\pi v_F^* \tilde\b^* \over  2 \L_0^2}\sum_{n=0}^{2}a_n^L \sum_{j} \hat F_{n}^{j\dagger} \hat F_n^{j}+\ldots,  \label{Eq:HLdec}
\ee
and coefficients $a_n^L$ are listed in Table~\ref{tab:SC_decomp}.

We see that the interaction mediated by the longitudinal phonons also favors $s$-wave pairing, hence, its only effect is to modify the expression for $T_c$ accordingly. The inclusion of the Coulomb interaction, on the other hand, may have more dramatic consequences, leading, under certain conditions, to the $p$-wave superconductivity in covalent crystals.

 \subsubsection{Possibility of $p$-wave pairing \label{Sec:p-wave}}

 To demonstrate how the Coulomb repulsion may result in the $p$-wave superconductivity, we generalize our analysis for the case of a finite Dirac mass $m$ in Eq.~\eqref{Spsi}. Again, we focus on the regime with $m/v_F,\, k_F \lesssim \Lambda_r$, so the RG flow is not affected by the nonzero mass/Fermi energy, and stops at the same scale $\Lambda_r$, while the ratio $m/v_F k_F$  can be arbitrary.

In case of a finite mass, the eigenvectors in the MCBB~\eqref{Eq:MCBB} are generalized as
\be
b_1(\bs k) = \left( \begin{array}{c}  \beta_+ - \beta_-\hat k_z \\ -  \beta_- \hat k_+ \\ \beta_+ + \beta_-\hat k_z \\ \beta_- \hat k_+    \end{array}  \right), \, b_2(\bs k) = \left( \begin{array}{c} -\beta_- \hat k_- \\  \beta_+ + \beta_- \hat k_z \\ \beta_-\hat k_- \\ \beta_+ - \beta_-\hat k_z  \end{array}  \right), \label{Eq:MCBB1}
\ee
where we only consider states at the Fermi surface, $|\bs k| = k_F$, and defined  $\beta_{\pm} = (1/2) \sqrt{1 \pm m/\ve_F}$ with $\ve_F = \sqrt{m^2 + v_F^2 k_F^2}$. Equation~\eqref{Eq:MCBB1} also assumes the Fermi energy inside the electron band, while the expression for the opposite case is obtained by the substitution $\beta_- \to \beta_+, \, \beta_+ \to - \beta_-$.  Hereafter, all quantities entering the equations (e.g., Fermi velocity $v_F$, Fermi momentum $k_F$, mass $m$, Fermi energy $\ve_F,$ Thomas-Fermi vector $q_{TF},$ or density of states $\nu$) are meant to be taken at the RG scale $l_r$ (which corresponds to $\Lambda_r$ in momentum space), and we suppress index ${}^*$ for brevity, unless  otherwise  specified.

The effect of a finite mass on the phonon-mediated part of the interaction is rather simple: It results in the extra prefactor $v_F^2  k_F^2 / \ve_F^2$ in Eqs.~(\ref{HFE}) and (\ref{HFEL}). As a result, all coefficient $a_n^T$ in Eq.~\eqref{Eq:HBCS}  should be replaced by $a_n^T \to (v_F^2  k_F^2 / \ve_F^2) a_n^T$ (and analogously for all coefficients $a_n^L$ in Eq.~(\ref{Eq:HLdec})). Finally, the density of states $\nu$ in Eqs.~\eqref{Eq:Tc0} and~\eqref{Eq:Tcp} should be modified according to  $\nu=  \ve_F k_F /2 \pi^2 v_F^2$.

The decomposition of the Coulomb repulsion is more subtle. Because of its long-range nature, the momentum dependence of the interaction must also be taken into account. Taking the simple Thomas-Fermi approximation and projecting onto the MCBB, we find:

\begin{widetext}
\begin{align}
\mc H_{C} = 8 \pi \alpha^* v_F  \sum_{\bs k,\bs p, \bs q}\frac1{q^2 + q_{TF}^2}
&\left\{c_{\bs k + \bs q}^\dag\left[ (\beta_+^2 + \beta_-^2(\widehat{\bs k +\bs q} \cdot \widehat{\bs k })) + i \beta_-^2 \widehat{\bs k +\bs q} \times \widehat{\bs k }\cdot \bs \s\right] c_{\bs k}\right\} \times \nn \\ &\left\{c_{\bs k'- \bs q}^\dag\left[ (\beta_+^2 + \beta_-^2(\widehat{\bs k'-\bs q} \cdot \widehat{\bs k' })) + i \beta_-^2 \widehat{\bs k'-\bs q} \times \widehat{\bs k' }\cdot \bs \s\right] c_{\bs k'}\right\},\label{HC}
\end{align}
\end{widetext}
where {$q_{TF}^2 = 8 \pi N \alpha^*  \nu v_F $ } is the square of the Thomas-Fermi wavevector, and $\alpha^*$ is given by
\be
\alpha^{*} \equiv  \alpha(l_r) = {\alpha_0 \over 1+{2(N+1)\over 3\pi} \alpha_0 \log{\L_0\over \L_r}}.
 \ee
Again, all quantities entering $q_{TF}$ here are taken at the RG scale $l_r.$

Focusing on the states at the Fermi surface only, we decompose the Coulomb interaction~\eqref{HC} into the pairing channels:
\be
\mc H_{C} \approx  {\pi  \alpha^* v_F\over k_F^2} \sum_{n=0}^{2}f_n \left(\frac{q_{TF}}{k_F}\right) \sum_{j} \hat F_{n}^{j\dagger} \hat F_n^{j}+\ldots, \label{Eq:Hcdec}
\ee
where the ellipsis stands for the terms with $J>1$ which we neglect here. The expression for functions $f_n(x)$ are rather cumbersome and presented in Appendix~\ref{App:Decomposition}.

Summing up contributions  from the transverse and longitudinal phonon modes, Eqs.~\eqref{Eq:HBCS} and~\eqref{Eq:HLdec},  and direct Coulomb repulsion~\eqref{Eq:Hcdec}, the decomposition of the total effective electron-electron interaction into the pairing channels has form

\begin{align}
\mc H_{FE}^T + \mc H_{FE}^L + \mc H_{C} \approx  {\pi  v_F} \sum_{n=0}^{2} \left[ -a_n^T\frac{\beta^* v_F^2 k_F^2}{ 2 \Lambda_0^2 \ve_F^2}- \right. \nn \\ \left. - a_n^L\frac{\tilde \beta^* v_F^2 k_F^2}{ 2 \Lambda_0^2 \ve_F^2} +  \frac{\alpha^*}{k_F^2}f_n \left(\frac{q_{TF}}{k_F}\right)\right] \sum_{j} \hat F_{n}^{j\dagger} \hat F_n^{j}+\ldots, \label{Eq:HBCScov}
\end{align}
where coefficients $a_n^T$ and $a_n^L$ are presented in Table~\ref{tab:SC_decomp}, and functions $f_n(x)$ are listed in Appendix~\ref{App:Decomposition}. Equation~\eqref{Eq:HBCScov} is a direct generalization of Eq.~\eqref{Eq:HBCS} for the case of covalent crystals and finite Dirac mass. The expressions for $T_c$ in the attractive pairing channels can also be easily generalized for this case.

In general, functions $f_n(x)$ have rather complicated form. However, to demonstrate how the $p$-wave superconductivity may appear, it is sufficient to consider the limit of a very low density, $v_F k_F \ll \ve_F = \sqrt{m^2 + v_F^2 k_F^2}.$ Assuming further that $\alpha^*$ is not too small, we find
{
\begin{align}
f_0\left(\frac{q_{TF}}{k_F}\right) &\approx \frac{\pi}{4N \alpha^*} \frac{v_F k_F}{\ve_F}, \nn  \\
f_1\left(\frac{q_{TF}}{k_F}\right) &\approx \frac{\pi^2}{24 N^2 \alpha^{*2}} \left(\frac{v_F k_F}{\ve_F}\right)^2, \nn  \\ f_2\left(\frac{q_{TF}}{k_F}\right) &\approx \frac{\pi^2}{16 N^2 \alpha^{*2}} \left(\frac{v_F k_F}{\ve_F}\right)^2. \label{Eq:fi}
\end{align}}

It is clear from the above expression that, as long as $\alpha^* \ve_F \gg v_F k_F,$ the $s$-wave channel is much more suppressed by the Coulomb repulsion than the $p$-wave channel, $f_0 \gg f_{1,2}$. We further assume that the coupling constants $\beta^*$ and $\tilde \beta^*$ are renormalized significantly enough, such that $\beta^* \sim 1$ and $\tilde \beta^* \approx 0$. It means that the system is on the verge of entering the strong-coupling regime, while the contribution from the interaction mediated by the longitudinal phonons can be neglected. Then, the ratio between the phonon-mediated attraction and the Coulomb repulsion in the $s$-wave channel can be rudely estimated as

{
\be
\left|\mc H_{FE}^{T(0)}/\mc H_{C}^{(0)} \right|\sim \beta^* N \left(\frac{v_F k_F}{\ve_F}\right) \left( \frac{k_F}{\Lambda_0}\right)^2 \ll1.
\ee
}
We see that, because of the small factor $v_F k_F / \ve_F \ll 1$, the Coulomb repulsion significantly exceeds the attraction due to phonons, thus completely suppressing superconductivity in this channel.

On the other hand, the analogous estimate for the vector-type $p$-wave pairing channel $F_2^i$, which, according to Table~\ref{tab:SC_decomp}, is also attractive if only the transverse phonons are considered, gives

\be
\left|\mc H_{FE}^{T(2)}/\mc H_{C}^{(2)} \right|\sim \alpha^* \beta^* N^2 \left(\frac{k_F(l_r)}{\Lambda_0}\right)^2,  \label{Eq:ratio}
\ee
where we explicitly restored the argument $l_r$. We see that, unlike the $s$-wave, the above expression does not have the smallness $v_F k_F /\ve_F$. Consequently, assuming that $\alpha^*,\, \beta^* \sim 1$, the ratio~\eqref{Eq:ratio} can be of the order of 1 provided the smallness $k_F(l_r) / \Lambda_0 = k_F / \Lambda_r \sim k_F c_{T0}/ \omega_{T0}$ is compensated by a large numerical prefactor and a large number of Dirac cones $N$.

The prerequisites for the $p$-wave superconductivity in the described above scenario impose a lot of constraints on the parameters entering the problem.  It is important to note that the spin-orbit effects are  suppressed in the limit $v_F k_F \ll \ve_F = \sqrt{m^2+v_F^2 k_F^2}$ (the dispersion becomes effectively Schr\"{o}dinger-like), which  results in the small  prefactor $v_F^2  k_F^2 / \ve_F^2$ in the effective interactions~(\ref{HFE}) and (\ref{HFEL}).  Because of this additional density dependence in the exponent that dictates $T_c$,  the TO phonon-mediated mechanism is not  parametrically greater than the standard acoustic phonons mechanism. Thus, the two mechanisms must be numerically compared to dictate who gives a larger $T_c$. Nonetheless, for low-density systems, where the density of states is small, they are both expected to give very small transition temperatures.

\section{Conclusions \label{Sec:conclusions}}
We have studied the ferroelectric quantum critical point in three-dimensional low-density Dirac materials. We  derived a general low-energy effective field theory that includes the interaction between soft phonon modes and electrons, as well as Coulomb repulsion. We showed that the dominant interaction between electrons is mediated by the transverse phonon mode, while the Coulomb repulsion is screened by the lattice. Using RG analysis, we demonstrated that the effective low-energy theory flows to a regime with strong electron-phonon coupling. For comparison, we performed similar analysis for covalent crystals, where the ``ferroelectric'' transition implies the breaking of inversion symmetry of the lattice without generating an electrical polarization. While the main results in this case are qualitatively the same, the screening of the Coulomb repulsion is much weaker in covalent crystals because of the lack of lattice polarization. We further demonstrated that the proximity to the FE critical point significantly enhances superconductivity. Finally, we showed how the interplay between phonon-mediated attraction and Coulomb repulsion in covalent crystals can, in principle, lead to $p$-wave superconductivity.

It is interesting that the problem of a Dirac semimetal undergoing a ferroelectric transition generically flows to strong coupling even in the absence of a finite Fermi surface. This problem can be addressed using the determinental quantum Monte Carlo method, since it does not suffer from a sign problem.  It will be informative to study the fate of the system in the strong coupling limit; for example, whether the superconducting transition survives, or it is destroyed by strong critical fluctuations.

We also expect that the strong-coupling regime close to the critical point will have experimental consequences. For example, high-accuracy measurements of the phonon dispersion may reveal a $2k_F$ Kohn anomaly.
Furthermore, strong coupling will significantly enhance electron scattering, leading to a strong temperature-dependent resistivity.

Considering our results in a broader context, we expect this mechanism to be relevant to all low-density superconductors that possess a near crossing of conduction and valence bands. As explained in the introduction, this feature applies to almost all low-density  superconductors, including bismuth, YPtBi, PbTe, SnTe, Sr$_x$Bi$_2$Se$_3$, Ge and Sr$_{3-x}$SnO. Of particular interest is the quadratic-band-touching semimetal YPtBi, where a similar RG analysis can lead to nontrivial fixed points. Finally, we suggest that the mechanism considered in this work may be relevant to the high-$T_c$ low-density superconductor FeSe, which also possesses a Dirac-like dispersion close to the $M$-points in the Brillouin zone.

\section{Acknowledgments}
We thank Patrick Lee, Liang Fu, Hiroki Isobe, Meng Cheng, and Kamran Behnia for illuminating discussions. V. K. is supported by DOE Office of Basic Energy Sciences, Division of Materials Sciences and Engineering under Award DE-SC0018945. Z. B. acknowledges support from the Pappalardo fellowship at MIT.

\bibliography{FE}

\appendix

\begin{widetext}

\section{One-loop diagrams} \label{app:diags}
Here we list the one-loop diagrams that contribute to our renormalization group equations. We start from the action for the covalent crystals  by including both the transverse and longitudinal optical phonon modes. The result for the polar case can be deduced by restricting to the diagrams with only transverse phonons (formally just take $e\rightarrow 0,\,  c_L\rightarrow \infty$ limit).
\begin{align}
\mc S = \int d^4 x \left\{ \sum_{n=1}^N\bar\psi_n \left[ Z_\psi \g_0 \partial_0  + v_F \g_j \partial_j \right] \psi_n + {1\over 2}u_j\left[ \left(- Z_u^2\partial_0^2+\w_T^2 \right)\d_{jl} -c_T^2\left(\nabla^2\d_{jl}-\partial_j\partial_l  \right)-c_L^2\partial_j\partial_l \right]u_l \right. \nonumber \\ \left.
+ V \left(u_j u_j \right)^2 +{\ve_\infty \over 8\pi} \left( \partial_j \phi\right)^2 +ie \sum_{n=1}^N\bar\psi_n \g^0 \psi_n \phi+\l \sum_{n=1}^N \bar \psi_{n} \g_0\g_j \psi_{n}\,u_j \right\}.
\end{align}
In this paper, we use the standard momentum-shell RG scheme, implying integrating out an infinitesimal momentum shell $\Lambda_0 e^{-\delta l} < q < \Lambda_0$ and all frequencies $-\infty < \omega < \infty$ at every RG step.

{\it Fermion self-energy --}
Both the optical phonons and the Coulomb field contribute to the fermion self-energy. The contributions from the phonon fields come from both the transverse phonon field and the longitudinal phonon field.
\begin{align}
\label{eq:FSE1}
\Sigma_\psi^u(k)&=\begin{gathered}
\includegraphics{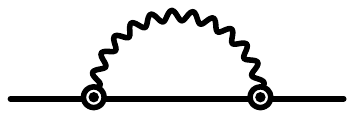}
\end{gathered}=
\Sigma_\psi^{uT}(k) +\Sigma_\psi^{uL}(k) \\ \nn &= -{iZ_u\l^2 dl \over 2 \pi^2 c_T\left( v_F Z_u + c_T Z_\psi\right)^2} \left( Z_{\psi}\omega \g_0+ {v_F\over 3} \bs k\cdot \bs{\g}\right) -{iZ_u\l^2 dl \over 4 \pi^2 c_L\left( v_F Z_u + c_L Z_\psi\right)^2} \left[ Z_{\psi}\omega \g_0- \left( 1+{2 c_L Z_\psi \over 3 v_F Z_u}\right){v_F} \bs k \cdot \bs \g\right].
\end{align}
The self-energy from the Coulomb interaction reads
\begin{align}
\Sigma_\psi^\phi(k) =\begin{gathered}
\includegraphics{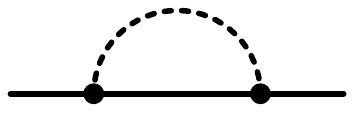}
\end{gathered}
=- {2e^2\,dl \over 3 \pi
\ve Z_\psi v_F } v_F \bs{k}\cdot\bs{\g}.
\end{align}

{\it Optical phonon self-energy --}
The polarization of the phonon field given by the fermion bubble diagram is:
\begin{align}
\Pi_u^\psi(q)= \begin{gathered}
\includegraphics{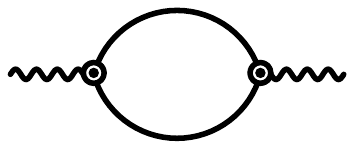}
\end{gathered}
 ={N\lambda ^2  dl\over 3\pi^2Z_\psi  v_F}\left[ {\L_0^2} \d_{jl}- {Z_\psi^2 \w^2\over  4v_F^2}\d_{jl}+{\bs q^2\over 4 }\left(\d_{jl} -2\hat q_j \hat q_l \right)  \right].
\end{align}

Interestingly, we notice that the fermion bubble diagram renormalizes the velocities of the transverse mode and the longitudinal mode in opposite ways. In this sense, the fermions can make the two modes very different. When considering the ionic crystal case, we simply omit the one-loop renormalization of the longitudinal mode,  since it has a large gap.

Additionally, the self-interaction of the phonon field also generates a self-energy correction, which is given by
\be
\Pi_u^V(k)  = \begin{gathered}
\includegraphics{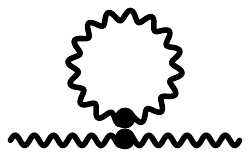}
\end{gathered}+ \begin{gathered}
\includegraphics{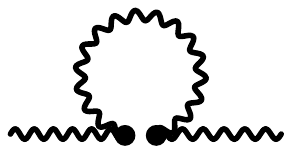}
\end{gathered}=-{5 V \Lambda_0^2 dl \over 3\pi^2}\left({2\over c_T}+{1\over c_L}\right).
\ee

{\it Coulomb field self-energy --}
\begin{align}
\Pi_\phi(q) =
 \begin{gathered}
\includegraphics{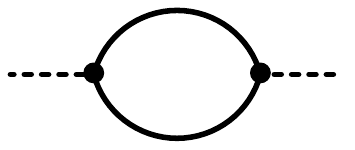}
\end{gathered}
=-{Ne^2 \, dl \over 6 \pi^2 Z_\psi v_F}\bs q^2.
\end{align}
{\it Electron-phonon vertex correction --}
Each vertex correction has two contributions, one where the boson exchanged in the loop is the same boson of the vertex and one where it is the other bosonic field. For example, the vertex correction to the electron-phonon coupling is given by
\begin{align}
\G_{u} &= \begin{gathered}
\includegraphics{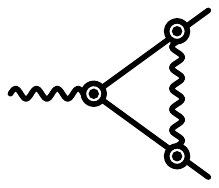}
\end{gathered}+ \begin{gathered}
\includegraphics{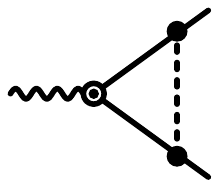}
\end{gathered} \\
&= -\lambda{(-  \lambda)^2 \over(2\pi)^4}\int d^4 p{\d_{jn}-\hat p _j \hat p_n\over (Z_u p_0)^2+(c_T\bs p)^2}\g_0\g_j {-i\over Z_\psi p_0\g_0+v_F \bs p \cdot \bs \g}  \g_0\g_l {-i\over Z_\psi p_0\g_0+v_F \bs p \cdot \bs \g}\g_0\g_n \nn \\
& -\lambda{(-  \lambda)^2 \over(2\pi)^4}\int d^4 p{\hat p _j \hat p_n\over (Z_u p_0)^2+(c_L\bs p)^2}\g_0\g_j {-i\over Z_\psi p_0\g_0+v_F \bs p \cdot \bs \g}  \g_0\g_l {-i\over Z_\psi p_0\g_0+v_F \bs p \cdot \bs \g}\g_0\g_n \nn \\
 &-\lambda{(-  ie)^2 \over(2\pi)^4}\int d^4 p{4\pi \over \ve_{\infty} \bs p^2}\g_0 {-i\over Z_\psi p_0\g_0+v_F \bs p \cdot \bs \g}  \g_0\g_l {-i\over Z_\psi p_0\g_0+v_F \bs p \cdot \bs \g}\g_0 \nn \\
&= -\lambda\g_0\g_l  \left[ {\l^2   \over 6\pi^2 c_T v_F \left(v_F Z_u + c_T Z_\psi\right)}-{\l^2 \left( 1+ {2 v_F Z_u\over v_F Z_u +c_L Z_\psi} \right)  \over 12\pi^2 c_L v_F \left(v_F Z_u + c_L Z_\psi\right)} + {e^2 \over 3 \pi v_F Z_\psi \ve_{\infty}} \right]\,dl. \nn
\end{align}

{\it Coulomb vertex correction --} The correction to the Coulomb vertex equals:
\begin{align}
\G_{\phi} &=
\begin{gathered}
\includegraphics{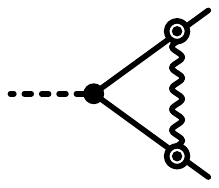}
\end{gathered} \\
&= -ie{(-  \lambda)^2 \over(2\pi)^4}\int d^4 p{\d_{jn}-\hat p _j \hat p_n\over (Z_u p_0)^2+(c_T\bs p)^2}\g_0\g_j {-i\over Z_\psi p_0\g_0+v_F \bs p \cdot \bs \g}  \g_0 {-i\over Z_\psi p_0\g_0+v_F \bs p \cdot \bs \g}\g_0\g_n \nn \\
& -ie{(-  \lambda)^2 \over(2\pi)^4}\int d^4 p{\hat p _j \hat p_n\over (Z_u p_0)^2+(c_L\bs p)^2}\g_0\g_j {-i\over Z_\psi p_0\g_0+v_F \bs p \cdot \bs \g}  \g_0 {-i\over Z_\psi p_0\g_0+v_F \bs p \cdot \bs \g}\g_0\g_n \nn \\
&= -ie\g_0  \left[ {\l^2 Z_u   \over 2\pi^2 c_T  \left(v_F Z_u + c_T Z_\psi\right)^2}+{\l^2 Z_u  \over 4\pi^2 c_L  \left(v_F Z_u + c_L Z_\psi\right)^2}\right]\,dl. \nn
\end{align}
Note that the one-loop $\sim e^3$ vertex correction vanishes in case of the instantaneous Coulomb interaction.

{\it Phonon interaction vertex correction--}
Finally, the four-phonon vertex is renormalized by the phonon bubble diagrams and by the fermion box diagram. Over all, we find that the correction is given by
\begin{align}
\label{eq:V}
\Gamma_V &=\begin{gathered}
\includegraphics{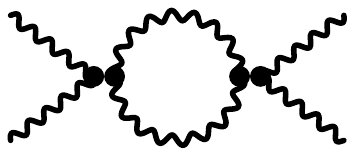}
\end{gathered}+\begin{gathered}
\includegraphics{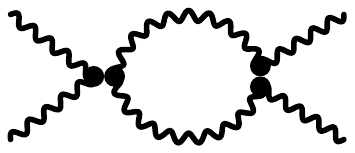}
\end{gathered}+\begin{gathered}
\includegraphics{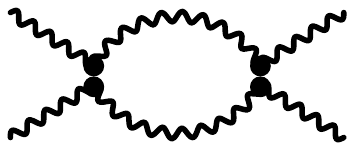}
\end{gathered}+\begin{gathered}
\includegraphics{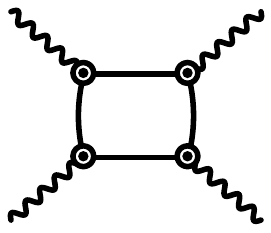}
\end{gathered}
\\ \nn
&=\left(\frac{17 V^2}{5\pi^2 Z_u c_T^3}+\frac{47 V^2}{30\pi^2 Z_u c_L^3}+{16 V^2\over 15\pi^2 Z_u c_Tc_L(c_T+c_L)} +\frac{\lambda^4 N}{24 \pi^2 Z_{\psi} v_F^3} \right)dl.
\end{align}

\section{RG equations for ionic crystals $(Q \ne 0)$}
\label{app:polar}
As explained in Section~\ref{sec:mu_ne_0}, in the ionic case the, dipolar interactions between lattice distortions generate a big mass $\omega_L$ for longitudinal phonons, which effectively screen the Coulomb repulsion between electrons. As a result, the Coulomb interaction and longitudinal phonons become irrelevant for our renormalization group study. Thus, the RG equations for the ionic case can be easily derived from the calculations in Appendix~\ref{app:diags} by setting $e = 0$ and $c_L \rightarrow \infty$. This procedure leads to the following RG equations

\begingroup
\allowdisplaybreaks
\begin{align}
&{d Z_{\psi} \over dl} = \left[2\eta_\psi - 2z - 3  +{\lambda^2 Z_u  \over 2 \pi^2 \mrm  c_T  (\mrm  c_T Z_{\psi} + v_F Z_u)^2} \right] Z_\psi \label{RGZpsi}\\
&{d v_F \over dl} =\left[2\eta_\psi - z - 4+{\lambda^2 Z_u  \over 6 \pi^2 \mrm  c_T  (\mrm  c_T Z_{\psi} + v_F Z_u)^2} \right]v_F\label{RGv} \\
&{d Z_{u}^2 \over dl} = (2 \eta_u -3z-3)Z_u^2+{N\lambda ^2 Z_\psi  \over 12\pi^2 v_F^3}\label{RGZu}\\
&{d  c_T^2 \over dl} =(2\eta_u-z-5) c_T^2-{N\lambda ^2  \over 12\pi^2Z_\psi  v_F}\label{RGuT}\\
&{d \omega_T^2 \over dl} =(2\eta_u -z-3)\omega_T^2-{N\lambda ^2  \L_0^2\over 3\pi^2Z_\psi  v_F} + {10 \L_0^2 V \over 3 \pi^2 c_T Z_u}\label{RGOm}\\
&{d \lambda \over dl} =(2\eta_\psi+\eta_u -2z-6)\l + {\l^3 \over 6\pi^2c_T v_F \left(v_F Z_u + c_T Z_\psi \right)}\label{RGlam}\\
&{d V \over dl} =(4 \eta_u -3z-9)V-{17 V^2 \over 5 \pi^2 c_T^3 Z_u}- \frac{\lambda^4 N}{24 \pi^2 Z_{\psi} v_F^3}
\end{align}
\endgroup

\section{RG equations for covalent crystals $(Q = 0)$}
\label{app:nonpolar}
The RG equations for the covalent case can be readily obtained from Eqs. \eqref{eq:FSE1}-\eqref{eq:V}:

\begingroup
\allowdisplaybreaks
\begin{align}
&{d Z_{\psi} \over dl} = \left[2\eta_\psi - 2z - 3  +{\lambda^2 Z_u  \over 2 \pi^2 \mrm  c_T  (\mrm  c_T Z_{\psi} + v_F Z_u)^2}+{\lambda^2  Z_u \over 4 \pi^2 \mrm  c_L  (\mrm  c_L Z_{\psi} + v_F Z_u)^2} \right] Z_\psi \label{RGZpsi2}\\
&{d v_F \over dl} =\left[2\eta_\psi - z - 4+{\lambda^2 Z_u  \over 6 \pi^2 \mrm  c_T  (\mrm  c_T Z_{\psi} + v_F Z_u)^2}-{\lambda^2 Z_u\left(3 v_F Z_u+ 2 c_L Z_\psi \right) \over 12 \pi^2 \mrm  c_L v_F Z_u  (\mrm  c_L Z_{\psi} + v_F Z_u)^2}+{2e^2 \over 3 \pi
\ve_{\infty} Z_\psi v_F } \right]v_F\label{RGv2} \\
&{d Z_{u}^2 \over dl} = (2 \eta_u -3z-3)Z_u^2+{N\lambda ^2 Z_\psi  \over 12\pi^2 v_F^3}\label{RGZu2}\\
&{d  c_T^2 \over dl} =(2\eta_u-z-5)  c_T^2-{N\lambda ^2  \over 12\pi^2Z_\psi  v_F}\label{RGuT2}\\
&{d  c_L^2 \over dl} =(2\eta_u-z-5)  c_L^2+{N\lambda ^2  \over 12\pi^2Z_\psi  v_F}\label{RGuL2}\\
&{d \omega_T^2 \over dl} =(2\eta_u -z-3)\omega_T^2-{N\lambda ^2  \L_0^2\over 3\pi^2 Z_\psi  v_F} + {5 \L_0^2 V \over 3 \pi^2 Z_u}\left({2 \over c_T}+{1\over c_L}\right)
\label{RGOmT2}\\
&{d  \ve_{\infty} \over dl} =(2\eta_\phi-z-5)  \ve_{\infty}+{2Ne^2 \over 3 \pi Z_\psi v_F}\label{RGeps2}\\
&{d \lambda \over dl} =(2\eta_\psi+\eta_u -2z-6)\l + {\l^3 \over 6\pi^2c_T v_F \left(v_F Z_u + c_T Z_\psi \right)}-{\l^3 \left( 3 v_F Z_u +c _L Z_\psi\right)  \over 12\pi^2 c_L v_F \left(v_F Z_u + c_L Z_\psi\right)^2} + {\l e^2 \over 3 \pi v_F Z_\psi \ve_{\infty}} \label{RGlam2}\\
&{d e \over dl} =(2\eta_\psi+\eta_\phi -2z-6)e +{e\l^2  Z_u \over 2\pi^2 c_T  \left(v_F Z_u + c_T Z_\psi\right)^2}+{e\l^2  Z_u \over 4\pi^2 c_L  \left(v_F Z_u + c_L Z_\psi\right)^2}\label{RGe2}\\
&{d V \over dl} =(4 \eta_u -3z-9)V - \frac{\lambda^4 N}{24 \pi^2 Z_{\psi} v_F^3} - \frac{17 V^2}{5\pi^2 Z_u c_T^3} - \frac{47 V^2}{30\pi^2 Z_u c_L^3}-{16 V^2\over 15\pi^2 Z_u c_Tc_L(c_T+c_L)}
\end{align}
\endgroup
It is worth noticing that the one-loop $\sim e^3$ correction to $e$ in Eq.~(\ref{RGe2}) vanishes in case of the instantaneous Coulomb interaction.

\section{Comments about crystal anisotropy \label{App:Anisotropy}}
In the analysis presented in the main text, we have considered a fully rotational invariant system. In a realistic  crystal, however, there are always anisotropies. In this Appendix, we will discuss such anisotropies in ionic crystals with cubic symmetry.

A cubic anisotropy has two important effects that are relevant to the flow of $\b$. First, the dispersion of the soft modes \eqref{Su} includes the anisotropy term
\[
\mc S_u^{a} = -\kappa\int d^4 q\, q_j^2 u_j^2.
\]
We neglect this term in what follows, i.e., we assume $\k = 0$. We also mention a recent comment where the effect of this term on the polarization of the ferroelectric modes was computed perturbatively \cite{ruhman-comment}.

The second effect of crystal anisotropy which we will consider appears in the electronic dispersion, which is relevant only for $N>1$. Let us consider $N=4$, where the four Dirac points occur on the boundary of the BZ at the $L$-points. In this case, the Dirac dispersion term Eq.~\eqref{Spsi} is modified according to~\cite{hsieh2012topological}

\be\label{Spsianiso}
\mc S_{\psi} = \sum_{n=1}^N\int d^4 x \,\bar\psi_n \left[\gamma_0 \partial_0+v_F \left(\gamma_x \partial_x+\gamma_y \partial_y \right) + v_z \g_z \partial_z  +m -  \gamma_0 \ve_F\right] \psi_n
\ee
where $z$-direction is defined differently for each Dirac point. It corresponds to the line connecting the $\G$-point to each of the $L$-points.
Additionally, the coupling of each of these Dirac points to the phonons Eq.~\eqref{Spsi-u} is also modified
\be\label{Spsiuaniso}
\mc S_{\psi u} = \sum_{n=1}^N\int d^4 x \, \bar \psi_n\left[ \l_z\g_0 \g_z\,u_z +\l \left(\g_0 \g_x\,u_x+\g_0 \g_y\,u_y\right)\right]\psi_n,
\ee
where $\l_z$ is the coupling to a distortion along the line and $\l$ is the coupling transverse to it.

\begin{figure}[t!]
 \begin{center}
    \includegraphics[width=.6\linewidth]{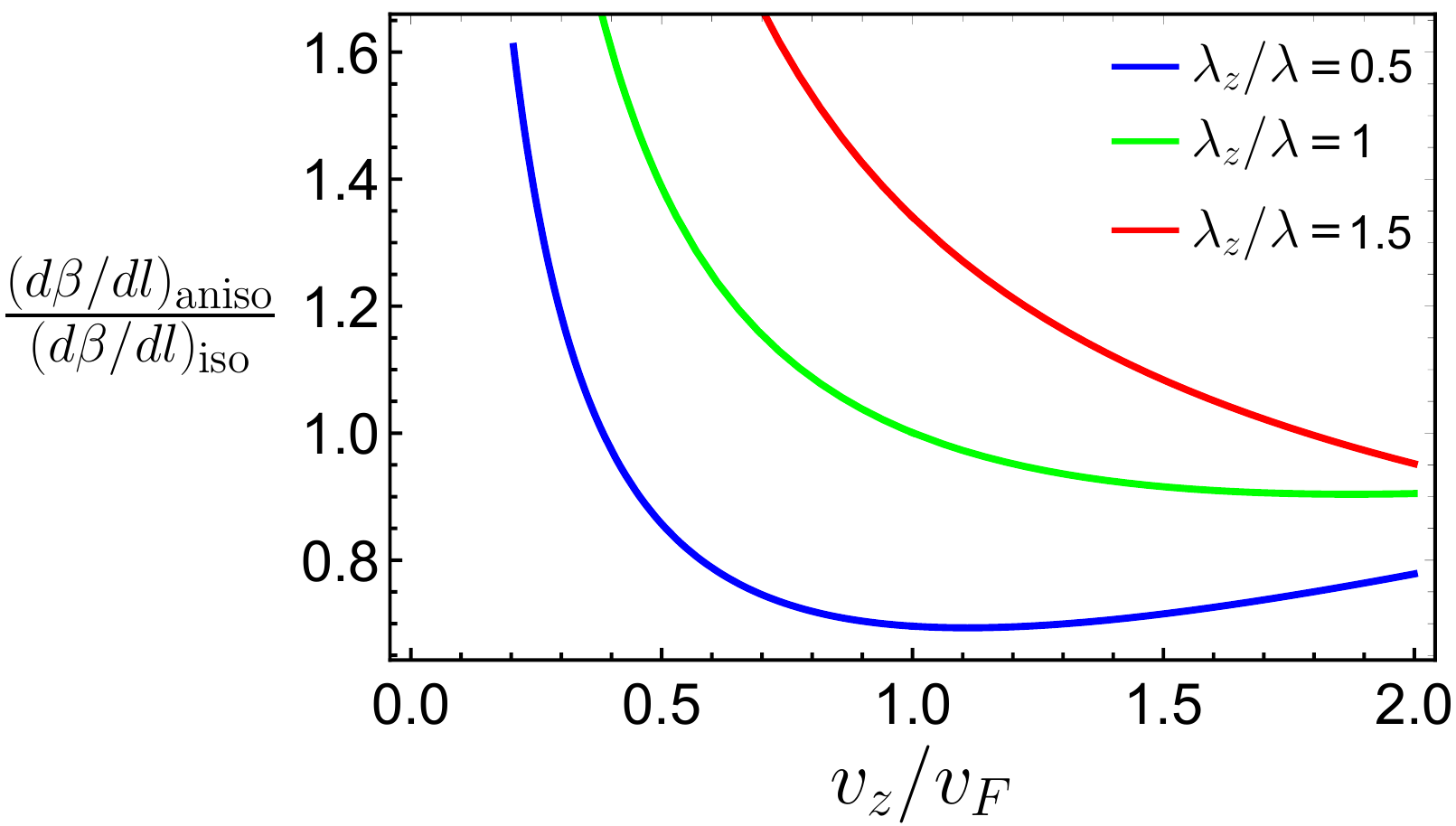}
 \end{center}
\caption{The $\mrm{\b}$-function for the flow of the dimensionless coupling constant $\b$ with the anisotropies, Eqs.~\eqref{Spsianiso}-\eqref{Spsiuaniso}, normalized by the ``isotropic'' $\mrm{\b}$-function given by Eq.~\eqref{RG_eqs_polar}, as a function of $v_z/v_F$ for three different values of $\l_z/\l$. }
 \label{fig:aniso}
\end{figure}

The RG procedure described in Sec.~\ref{sec:RG} can still be performed analytically, although the expressions become rather lengthy.
We also have two additional dimensionless parameters $ v_z / v_F$ and $ \l_z / \l$ that flow under RG.

The crucial point is that the main result of this paper, the flow of $\b$ towards strong coupling, remains unchanged. To demonstrate this point, we plot the $\b$-function for $\b$ in the anisotropic case normalized by the $\beta$-function for isotropic case, Eq.~\eqref{RG_eqs_polar}, with $\zeta_T = 0.1$ in Fig.~\ref{fig:aniso}. Here, the ratio is plotted for three different values of $\l_z/\l$ as a function of $v_z/v_F$. We find that in all cases the $\b$-function is positive. Also note that the regimes of $v_z<v_F$ and $v_z>v_F$ should not be considered on equal footing, since in the latter case the density of states at the Dirac point is enhanced while in the opposite limit it is decreased. Also, note that we did not perform a detailed study of the multidimensional flow in the four-dimensional space of all four parameters.

\section{Decomposition of the effective interaction into pairing channels \label{App:Decomposition}}

In this Appendix, we briefly outline the procedure for the decomposition of the effective interaction into the pairing channels. As an example, we consider the Coulomb interaction given by Eq.~\eqref{HC}, while the decomposition of the phonon-mediated interactions~\eqref{HFE} and~\eqref{HFEL} can be performed analogously. Considering only pairings with the zero total momentum, we find that Eq.~\eqref{HC} can be written as

\be
\mc H_{C} \approx  \sum_{\bs k,\bs p} V_{\alpha\beta \gamma \delta}( {\bs k}, {\bs p}) c^\dagger_{ {\bs p}\alpha}c^\dagger_{ {-\bs p}\beta}c_{ {-\bs k}\gamma}c_{ {\bs k}\delta},
\ee
where $V_{\alpha\beta \gamma \delta}({\bs k}, {\bs p})$ is given by

\begin{multline}
V_{\alpha\beta \gamma \delta}({\bs k}, {\bs p}) = 4 \pi \alpha^* v_F  \left\{  \frac{1}{({\bs p} - {\bs k})^2 + q_{TF}^2}
\left[ (\beta_+^2 + \beta_-^2(\widehat{\bs p} \cdot \widehat{\bs k })) + i \beta_-^2 \widehat{\bs p} \times \widehat{\bs k }\cdot \bs \s\right]_{\alpha \delta} \times\left[ (\beta_+^2 + \beta_-^2(\widehat{\bs p} \cdot \widehat{\bs k })) + i \beta_-^2 \widehat{\bs p} \times \widehat{\bs k }\cdot \bs \s\right]_{\beta \gamma} \right. - \\ - \left. \frac{1}{({\bs p} + {\bs k})^2 + q_{TF}^2}
\left[ (\beta_+^2 - \beta_-^2(\widehat{\bs p} \cdot \widehat{\bs k })) - i \beta_-^2 \widehat{\bs p} \times \widehat{\bs k }\cdot \bs \s\right]_{\alpha \gamma} \times\left[ (\beta_+^2 - \beta_-^2(\widehat{\bs p} \cdot \widehat{\bs k })) - i \beta_-^2 \widehat{\bs p} \times \widehat{\bs k }\cdot \bs \s\right]_{\beta \delta} \right\},
\end{multline}
and we only took into account states near Fermi surface, $|\bs k| \approx |\bs p| \approx k_F$. Then, this  interaction can be decomposed into the pairing channels according to

\be
V_{\alpha\beta \gamma \delta}({\bs k}, {\bs p}) =  {\pi  \alpha^* v_F\over k_F^2} \sum_{n=0}^{2}f_n \left(\frac{q_{TF}}{k_F}\right) \sum_{j}  (i F_{n}^{j}(\hat{\bs p}) \sigma_y)_{\alpha \beta} ( i F_n^{j}(\hat{\bs k}) \sigma_y)^{\dagger}_{\gamma \delta}+\ldots,
\ee
where form factors $F_n^j$ are defined in Table~\ref{tab:SC_decomp}. Next, multiplying this expression by $( i F_n^{j}(\hat{\bs k}) \sigma_y)_{\delta \gamma}$, performing the summation over spin indices $\gamma$ and $\delta$ using Fierz identities, and evaluating integral over $\hat{\bs k}$, we find

\begin{align}
f_0(x) &= (\beta_+^4 + \beta_-^4) \ln\frac{4+x^2}{x^2} + \beta_+^2 \beta_-^2 \left[ -4 + (2+x^2)\ln\frac{4+x^2}{x^2}  \right],  \\
f_1(x) &= 2\beta_+^2 \beta_-^2\ln\frac{4+x^2}{x^2}  + \frac{\beta_+^4 + \beta_-^4}{2}\left[ -4 + (2+x^2)\ln\frac{4+x^2}{x^2}  \right], \nonumber \\ f_2(x) &= \frac{3\beta_+^2 \beta_-^2}{4} \left[ -2(2+x^2) + \frac{4+(2+x^2)^2}2 \ln\frac{4+x^2}{x^2} \right] + \frac{3(\beta_+^4 + \beta_-^4)}2\left[ -2 + \frac{2+x^2}2\ln\frac{4+x^2}{x^2}  \right]. \nn
\end{align}
The asymptotic behavior of these expressions in the limit $k_F v_F \ll \ve_F = \sqrt{m^2 + k_F^2 v_F^2}$ (which corresponds to $x\gg1$) is presented in the main text, see Eq.~\eqref{Eq:fi}.

\end{widetext}
\end{document}